\documentclass[11pt,a4paper]{article}

\usepackage[utf8]{inputenc}
\usepackage[T1]{fontenc}
\usepackage{microtype}
\usepackage{graphicx}
\usepackage{xcolor}
\usepackage{hyperref}
\usepackage{geometry}
\usepackage{titlesec}
\usepackage{titling}
\usepackage{fancyhdr}
\usepackage{enumitem}
\usepackage{booktabs}
\usepackage{float}
\usepackage{listings}
\usepackage{fancyhdr}
\usepackage{algorithm}
\usepackage{algpseudocode}
\usepackage{listings,booktabs,enumitem}

\usepackage{lmodern}

\definecolor{primarycolor}{RGB}{0, 123, 255}  
\definecolor{secondarycolor}{RGB}{108, 117, 125}  
\definecolor{accentcolor}{RGB}{40, 167, 69}  

\hypersetup{
  colorlinks=true,
  linkcolor=primarycolor,
  filecolor=primarycolor,
  urlcolor=primarycolor,
  citecolor=primarycolor
}

\titleformat{\section}
  {\normalfont\Large\bfseries\color{primarycolor}}
  {\thesection}{1em}{}
\titleformat{\subsection}
  {\normalfont\large\bfseries\color{secondarycolor}}
  {\thesubsection}{1em}{}

\pretitle{\begin{center}\LARGE\bfseries\color{primarycolor}}
\posttitle{\end{center}\vspace{1em}}
\preauthor{\begin{center}\large}
\postauthor{\end{center}}
\predate{\begin{center}\large}
\postdate{\end{center}\vspace{2em}}

\begin{document}

\title{\textcolor{primarycolor}{Coral Protocol}\\
\Large\textcolor{secondarycolor}{Open Infrastructure Connecting\\The Internet of Agents}}
\author{Roman J. Georgio, Caelum Forder, Suman Deb, Andri Rahimov,\\Peter Carroll, \"{O}nder G\"{u}rcan*\\
\small{Contact: hello@coralprotocol.com | www.coralprotocol.org}}
\date{\today}
\maketitle

\begin{abstract}
Coral Protocol is an open and decentralized collaboration infrastructure that enables communication, coordination, trust and payments for The Internet of Agents. 
It addresses the growing need for interoperability in a world where organizations are deploying multiple specialized AI agents that must work together across domains and vendors.
As a foundational platform for multi-agent AI ecosystems, Coral establishes a common language and coordination framework allowing any agent to participate in complex coordinations with others. Its design emphasizes broad compatibility, security, and vendor neutrality, ensuring that agent interactions are efficient and trustworthy. 
In particular, Coral introduces standardized messaging formats for agent communication, a modular coordination mechanism for orchestrating multi-agent tasks, and secure team formation capabilities for dynamically assembling trusted groups of agents. Together, these innovations position Coral Protocol as a cornerstone of the emerging “Internet of Agents,” unlocking new levels of automation, collective intelligence, and business value through open agent collaboration.
\end{abstract}

\newpage
\tableofcontents
\newpage

\section{Introduction}

In recent years, the AI landscape has begun shifting from standalone systems toward networks of specialized agents working in concert. Advanced language models, decision-making bots, and domain-specific AI services are increasingly expected to interact with one another to tackle complex tasks that no single system can handle alone. This transition to multi-agent AI ecosystems promises significant gains in efficiency and capability, but it also raises a critical challenge: ensuring that these diverse agents can communicate and coordinate effectively across different platforms and organizations.

Today, efforts to connect AI agents remain fragmented. Various groups have introduced their own agent-to-agent frameworks—ranging from proprietary enterprise solutions to open-source projects—yet none has emerged as a universal standard. For example, major cloud providers have launched protocols like Google's Agent2Agent (A2A) to facilitate cross-platform agent interaction \cite{Surapaneni2025}, while open consortia such as Cisco’s AGNTCY initiative are pursuing an “Internet of Agents” vision for interoperable agent collaboration \cite{Sheth2025AGNTCY}. Similarly, researchers at MIT have introduced NANDA, a decentralized agent framework focused on robust coordination and composability across AI systems\footnote{NANDA: The Internet of AI Agents, \url{https://nanda.media.mit.edu}, accessed on April 18, 2025.}. 
These initiatives highlight the strong demand for agent interoperability. However, they remain largely siloed, each addressing only parts of the problem within limited ecosystems. The result is a patchwork of incompatible approaches that hinders agents developed in different environments from truly working together.

Coral Protocol is designed to unify these disparate efforts by providing a common foundation for AI agent collaboration. Conceived as an open and vendor-neutral infrastructure, Coral offers a standardized way for agents—regardless of who built them or what frameworks they run on—to communicate, share knowledge, and coordinate tasks. By serving as a general-purpose lingua franca for agents, Coral breaks down integration silos and accelerates the emergence of robust multi-agent services. It builds upon lessons learned from earlier protocols but goes further, combining communication, coordination, and security capabilities into one cohesive platform for agent interaction.

Unlike its predecessors that often focused on either networking or task orchestration in isolation, Coral delivers a holistic solution for multi-agent interoperability. Its core capabilities can be summarized as follows:

\begin{itemize}
    \item \textbf{Structured Interaction Mediation}: Coral manages all communication between users and agents, as well as between agents themselves. Through persistent threads and mention-based targeting, it ensures that conversations remain organized, contextual, and efficient—without requiring agents to poll continuously or interpret unstructured input.
    \item \textbf{Dynamic Agent Discovery and Capability Registration}: Agents can advertise their capabilities and discover others through standardized mechanisms, allowing seamless composition of multi-agent coordinations without hardcoded integrations or platform-specific logic.
    \item \textbf{Secure Team Formation and Task Execution Coordination}: Coral enables the on-demand assembly of agent teams with authenticated identities, assigned roles, and controlled data access. Teams can collaboratively execute complex tasks while preserving privacy, trust, and auditability via a secure coordination layer and optional blockchain-based logging.
    \item \textbf{Modular Tool and Data Integration}: Via Coralizer modules, developers can onboard models, tools, and datasets—transforming them into "Coralized" agents accessible within the ecosystem. This modular onboarding allows for scalable growth of capabilities without compromising interoperability.
    \item \textbf{Built-in Economic Transactions}: The protocol natively supports payment flows through a secure payment service. Agents can be compensated for their contributions, enabling incentive-aligned marketplaces of AI services and autonomous microtransactions between agents.
    \item \textbf{Infrastructure-Agnostic Deployment}: Powered by MCP servers and a decentralized architecture, Coral agents can run across heterogeneous environments while exposing standardized interfaces. This ensures composability at internet scale.
\end{itemize}

By filling this critical infrastructure gap, Coral Protocol aims to catalyze a vibrant agent ecosystem in which intelligent services can freely cooperate at scale. In essence, Coral aspires to become the linchpin of an interoperable “agent-of-agents” environment, transforming today’s siloed AI systems into a harmonized network of agents that achieves outcomes no isolated system could accomplish on its own.

\newpage
\section{Enabling Concepts}

In this section, we review several key concepts that underpin Coral Protocol's vision of interoperable AI agents. We cover recent progress in large language models, autonomous AI agents and their use of tools, the emergence of model context protocols, and the advent of multi-agent AI systems. Together, these advances set the stage for why a unifying interoperability protocol is needed.

\subsection{Large Language Models (LLMs)}

Large Language Models (LLMs) are very large neural networks trained on massive text datasets to predict and generate text. Modern LLMs are built on the transformer architecture, which uses self-attention mechanisms to capture long-range dependencies in text \cite{vaswani2017attention}. Scaling up model size and training data has proven remarkably effective: for example, OpenAI's GPT-3 (2020) with 175 billion parameters demonstrated that increasing model scale dramatically improves performance on a wide range of tasks without task-specific training, enabling strong \textit{few-shot} learning abilities \cite{brown2020language}. This means GPT-3 could be prompted with a few examples and then perform tasks like translation or Q\&A at near state-of-the-art levels \cite{brown2020language}. Such emergent capabilities were a surprise and highlighted the potential of "foundation models" – general-purpose models that can be adapted to many applications \cite{bommasani2021foundation}.

Since 2020, LLM development has accelerated. Researchers introduced techniques to make LLMs more useful and aligned with human intentions. Notably, fine-tuning models with human feedback and instructions (often via reinforcement learning from human feedback) has led to more reliable and safer AI assistants \cite{ouyang2022training}. This approach was used to create \textit{InstructGPT} and ultimately ChatGPT in 2022, greatly improving the model’s ability to follow user instructions politely and correctly \cite{ouyang2022training}. The year 2023 saw the release of even more advanced LLMs such as GPT-4, which not only improved text understanding and reasoning but also introduced multimodal capabilities (accepting image inputs) \cite{Shen2023HuggingGPT}. Other organizations have developed competitive LLMs (e.g., Google’s PaLM and Meta’s LLaMA), including open-source models, expanding access to this technology. Today’s state-of-the-art LLMs exhibit strong language understanding, reasoning via chain-of-thought, and even basic tool use. They serve as the "brains" for many AI agent systems and are the foundational technology that Coral Protocol builds upon. Practical applications of LLMs now range from code generation and data analysis to powering chat assistants and domain-specific experts. The rapid post-2020 progress in LLM capabilities and availability has set the stage for autonomous AI agents that can leverage these models’ general intelligence.

\subsection{AI Agents}

In AI, an \textit{agent} refers to a system that perceives its environment and takes actions to achieve goals. The concept of intelligent agents has long been studied in classical AI \cite{tran2025multiagent}, but recent advances in LLMs have given rise to a new breed of AI agents endowed with powerful language and reasoning skills. These agents use LLMs as their core, enabling them to interpret instructions, plan actions, and carry out complex tasks autonomously. In essence, an AI agent combines an LLM’s general knowledge with a decision-making loop: it can observe some input (or environment state), reason about what to do (often internally via "chain-of-thought" text), and then act (produce outputs or manipulate tools).

A key development enabling modern AI agents was the realization that LLMs can be prompted not just to answer questions, but to \textbf{think} step-by-step and \textbf{act} in a task-oriented manner. For example, the ReAct framework \cite{yao2022react} showed that an LLM can intermix reasoning steps with actionable commands, using its internal chain-of-thought to decide which external action to take next \cite{yao2022react}. In the ReAct paradigm, the model generates \textit{reasoning traces} (e.g. exploring possible solutions in text) and \textit{action commands} (e.g. queries to a tool or environment) in an interleaved way \cite{yao2022react}. This synergy of reasoning and acting allows an agent to break down complex problems, leverage external information when needed, and adjust its plan based on the results of its actions. For instance, an LLM-based agent can iteratively decide to perform a web search, read the results, and then use that information to answer a hard question – effectively self-guiding its behavior.

Beyond research prototypes, practical AI agent frameworks have proliferated. Several 2023 projects (many open-source) demonstrated \textit{autonomous agents} powered by GPT-4 or similar models that can execute multi-step plans. Examples include systems like "AutoGPT" and "BabyAGI," which loop an LLM’s outputs back into itself to create continuous task planning and execution cycles. These agents maintain a form of working memory (often a summary of past steps or an external vector database) and can spawn sub-agents for subtasks. While often experimental, they illustrate the growing aspiration for \textbf{AI agents that can operate with minimal human intervention}, handling tasks like researching a topic, managing schedules, or even controlling a computer. In academia, researchers have explored agents that carry out extended interactions or exhibit human-like behavior. For example, Generative Agents \cite{park2023generative} populated a simulated world with multiple LLM-driven characters that \textbf{plan, remember, and interact} dynamically, producing believable human-like behaviors over long periods. Such experiments show that LLM agents can in principle manage complex goals and social interactions when given appropriate architectures for memory and planning.

The practical relevance of AI agents is significant: they have the potential to automate tasks that normally require human-like judgment or complex decision processes, from customer service chats to orchestrating business workflows. Companies are beginning to deploy autonomous agents in roles like scheduling, IT support (e.g. auto-triaging requests), and data monitoring. However, creating robust agent systems remains challenging. Agents need to ground their decisions in reliable data (to avoid purely "imagining" incorrect actions) and operate within constraints (to ensure safety and alignment with user intentions). These needs motivate the use of external tools and standardized protocols, as we discuss next. Overall, the convergence of LLMs with the agent paradigm has led to a flurry of progress post-2020 in what some call "agentic AI" – AI systems that proactively take initiative and perform extended tasks. This sets the stage for networks of such agents working together, which is exactly the scenario Coral Protocol targets.

\subsection{Tools as Extensions of AI Agents}

A remarkable aspect of human intelligence is the ability to use tools – instruments that extend our capabilities. Similarly, AI agents can greatly expand their competence by using external tools to complement the knowledge contained in their model parameters. In the context of LLM-based agents, "tools" usually refer to any external system the agent can invoke via an API or function call: examples include search engines, databases, calculators, code execution environments, or even other machine learning models. By using tools, an AI agent can obtain up-to-date information, perform precise computations, interact with the physical world (through APIs to devices or services), and generally overcome the static knowledge limitations of its trained model.

The idea of tool-use in AI gained traction as researchers observed that LLMs, while very knowledgeable, still have constraints (e.g. fixed training data cutoff, limited factual accuracy on niche or recent info, etc.). 
Giving an agent the ability to fetch information or execute code on the fly can address these gaps. One early demonstration was OpenAI's \textbf{WebGPT}, where an LLM was augmented with a web browsing capability to improve the factual accuracy of its answers \cite{Nakano2022WebGPT}. Around the same time, techniques like Program-Aided Language Modeling and others showed that even math problem-solving by an LLM could be improved by calling a calculator or Python interpreter for difficult calculations. 

Tool use became a prominent research area by 2022. \cite{yao2022react}'s \textbf{ReAct} paradigm (mentioned above) is one approach that naturally integrates tool calls into an agent’s reasoning process. Another notable work is \textbf{Toolformer} \cite{schick2023toolformer}, where the LLM was trained to decide \textit{when} to invoke APIs and how to incorporate the results into its text generation \cite{Qu2024}. Toolformer demonstrated that a language model can learn to use tools such as a dictionary, calculator, or search engine in a zero-shot manner, leading to higher accuracy on tasks requiring those tools. The practical upshot is that an LLM doesn’t need to internally solve everything – it can learn to delegate sub-tasks to a tool that’s better suited for it.

A major real-world milestone in this area was the introduction of \textbf{API calling and plugins} for ChatGPT (2023). OpenAI enabled a standardized way for the model to invoke external functions provided by developers (e.g. retrieve stock prices, book a calendar event, or run a code snippet). This concept, analogous to a plugin system for the AI, showed how a single AI agent could interact with many services safely through a defined interface. Microsoft’s \textit{HuggingGPT} project went a step further: it treated ChatGPT as a controller that orchestrates calls to numerous expert models on HuggingFace (for vision, speech, etc.), essentially using specialist AI models as tools to solve multi-modal tasks \cite{Shen2023HuggingGPT}. By parsing a user request and breaking it into steps handled by appropriate models, HuggingGPT achieved impressive results across language, vision, and audio tasks that no single model could handle alone.

In summary, tool integration has become an essential part of advanced AI agent design. It offers \textbf{practical benefits}: agents can access real-time information (via web or database queries), perform actions on behalf of users (e.g., send an email, control a smart home device), and tackle problems (like math or code execution) that are difficult to solve with pure neural reasoning. Post-2020, we’ve seen rapid progress in frameworks that make tool-use easier – from research prototypes to robust libraries (e.g., LangChain, which simplifies connecting LLMs to tools). However, these integrations have often been ad-hoc, with each system defining its own set of tools and APIs. This fragmentation has paved the way for efforts to standardize how agents discover and use tools, which is where the \textbf{Model Context Protocol} comes in.

\subsection{Model Context Protocol (MCP)}
\label{sec:Model-Context-Protocol-(MCP)}

As AI agents proliferate, each with potentially different developers and tool suites, a clear challenge arises: How can an agent easily access a wide range of tools and data sources, especially those it wasn't originally built to use? And conversely, how can developers expose new tools or datasets such that any compliant AI agent can utilize them without custom integration? The \textbf{Model Context Protocol (MCP)} is a recently introduced answer to these questions. MCP is an open standard (first released in late 2024) that defines a common interface for connecting AI models (or agents) with external resources, in a way that is general and vendor-agnostic ([Introducing the Model Context Protocol \cite{Anthropic2024MCP}.

In essence, MCP provides a \textit{universal language} for AI agents to request access to data or functionality, and for tools/servers to offer that access. Anthropic, the company behind the Claude LLM, spearheaded MCP to break down the "silos" that trap AI systems away from the data and tools they need \cite{Anthropic2024MCP}. Traditionally, if you wanted your AI agent to use a new database or API, you had to wire up a bespoke connector for that specific combination of agent and tool. This led to an $N \times M$ integration problem (with $N$ AI systems and $M$ tools all needing pairwise connectors), causing duplication of effort and inconsistent implementations.

Concretely, MCP defines a client–server model. The \textbf{AI agent or LLM} acts as a \textit{client} that can send requests (for data, or to invoke an operation) in a standardized format. A \textbf{tool or data source} (e.g., a database, a knowledge base, an email service, or a custom function) runs as an MCP \textit{server} which knows how to interpret those standardized requests and execute the appropriate action, then return results. The protocol covers how an agent can discover available tools and what their capabilities are, how to call those functions with the right parameters, and how to handle security/authentication. For instance, an MCP server might expose a "database.query" capability or a "calendar.scheduleMeeting" function. Any MCP-enabled agent can invoke these without needing bespoke code for that specific database or calendar system. The agent simply sees the abstract interface. This setup \textbf{decouples} AI models from tools: agents and tools can be developed independently so long as both conform to MCP. This is analogous to how web browsers and web servers interoperate via HTTP – a universal protocol.

MCP’s design was inspired in part by the success of the Language Server Protocol (LSP) in software development \cite{surapaneni2025A2A}. Just as LSP allows any code editor to interface with any programming language’s analysis engine through a common protocol, MCP aims to let any AI agent interface with any tool or context provider. The MCP specification covers various types of "context" that an agent might need. These include \textbf{resources} (documents or data that the model can read), \textbf{tools/functions} (operations the model can invoke), and even shared \textbf{prompts or templates} (reusable pieces of guidance for the model). By standardizing these, MCP makes it easier to share not only data, but also behaviors. For example, a company could develop an MCP server for their internal knowledge base and another for their CRM system; any MCP-compliant AI assistant (from any vendor) could then query those seamlessly. This universality promises to reduce re-implementation and enable richer \textbf{ecosystems} of AI capabilities. Indeed, early adopters of MCP have demonstrated scenarios like an AI coding assistant pulling info from a project’s GitHub repository and issue tracker via MCP, or a customer support agent retrieving user history from a CRM – all through the same protocol \cite{Anthropic2024MCP}.

It’s worth noting that MCP is one of a few efforts emerging to standardize AI-tool interactions. Around the same time, Google introduced an \textit{Agent-to-Agent (A2A)} communication protocol with similar goals of interoperability \cite{surapaneni2025A2A}. The momentum behind these initiatives signals a recognition that as we integrate AI into many applications, we need common \textit{interfaces} to avoid each AI system becoming an isolated silo. For the general tech community, MCP is significant because it could do for AI what APIs did for web services – provide a lingua franca enabling diverse systems to work together. In the context of Coral Protocol, MCP represents a foundational step towards an open infrastructure where specialized AI agents can share context and services. Coral builds upon this idea of interoperability, extending it to not just tool access but also agent-to-agent collaboration in a broader network (as we’ll see later). In short, MCP and similar standards pave the way for \textbf{plug-and-play AI components}: you can mix and match models, tools, and data sources with minimal friction, which is crucial for scalable multi-agent ecosystems.

\subsection{Multi AI Agent Systems}

Moving beyond single agents, the next frontier is systems composed of \textit{multiple AI agents} working in concert. Just as groups of humans can collaborate by dividing labor or bringing different expertise, multiple AI agents could, in theory, tackle complex tasks more effectively by communicating and specializing. Research in multi-agent systems is not new – it has existed for decades in fields like distributed AI and robotics. However, \textit{LLM-based} multi-agent systems have only recently become feasible and are now an exciting area of development \cite{tran2025multiagent}. The general idea is to have a collection of agents (each potentially with different roles, knowledge, or abilities) that interact with each other to solve problems or create richer simulations.

There are several motivations for using multiple AI agents together. One is \textbf{specialization}: one agent might be an expert in math, another in coding, another in interacting with humans. By having them communicate, each sub-problem can be handled by the best-suited agent. 
For example, in a software development assistant, one agent could generate code while another agent reviews it for errors – akin to a pair programming scenario. Another motivation is \textbf{parallelism}: agents can work on different subtasks simultaneously and then share results, speeding up complex tasks. 
Yet another is \textbf{emergent behavior}: sometimes groups of agents can exhibit intelligent behaviors that single agents cannot, by bouncing ideas off each other or negotiating. A recent study by Tran et al. (2025) notes that LLM-based multi-agent systems enable "groups of intelligent agents to coordinate and solve complex tasks collectively at scale, transitioning from isolated models to collaboration-centric approaches" \cite{tran2025multiagent}. In other words, we’re beginning to move from thinking about one AI in isolation to networks of AIs collaborating, which some see as steps toward an "artificial collective intelligence."

In practical terms, multi-agent systems could transform how we use AI in large organizations or even in daily life. We might have an ensemble of specialized agents: one monitors news and data feeds, another manages your personal schedule, another handles creative brainstorming for your projects, and they talk to each other when needed. Enterprises are looking at multi-agent ecosystems where, say, a finance analysis agent, a marketing data agent, and an IT automation agent could interoperate to drive business processes end-to-end. Google’s recent announcement of the A2A (Agent-to-Agent) protocol highlights this vision: it emphasizes enabling agents "to collaborate in a dynamic, multi-agent ecosystem across siloed data systems and applications" and to interoperate even if built by different vendors \cite{surapaneni2025A2A}. In such an ecosystem, one agent could call upon another as a tool (e.g., a customer support agent hands off a complex technical query to a troubleshooting agent), or they might form a sequence (output of one is input to another). Realizing this smoothly will require common standards for agent communication and trust – precisely the type of infrastructure Coral Protocol aims to provide.

\subsection{Multi AI Agent Collaboration}

Several proof-of-concept systems have explored multi-agent collaboration. 
The CAMEL framework (2023) showed that two LLM agents can role-play as a "user" and an "assistant" to iteratively solve a task together, essentially allowing the AI to have a conversation with itself from different perspectives to refine a solution. Microsoft’s \textbf{AutoGen} \cite{Wu2024AutoGen} provides a programming framework for composing *conversational* interactions among multiple agents (and humans), so that one can build execution plans where agents ask each other for help \cite{Wu2024AutoGen}. 
For instance, one agent could be tasked with decomposing a problem into steps, then delegating each step to other specialized agents, and finally aggregating the results. Experiments with AutoGen found that such setups can handle complex queries in domains like coding or decision-making more effectively than a single agent working alone \cite{Wu2024AutoGen}. 
Another striking example is the \textit{Generative Agents} simulation mentioned earlier, where 25 agents inhabited a small virtual town – here the focus was on agents having social interactions with each other (e.g. sharing information, forming plans to organize a party) and it demonstrated that coherent multi-agent dynamics can emerge from relatively simple principles plus consistent language-model reasoning \cite{sumers2023cognitive}.

That said, coordinating multiple autonomous agents also introduces challenges. Without careful design, multiple agents could talk in circles, misinform each other, or work at cross purposes. There are open questions in research about \textbf{coordination strategies} (how do agents reach consensus or allocate tasks among themselves?) and \textbf{communication protocols} (what language or format should agents use to exchange information efficiently and without ambiguity?). Earlier multi-agent systems research introduced languages like KQML and frameworks like FIPA-ACL for agent communication, but those were largely pre-LLM and used in constrained settings. Now, with language-capable agents, one straightforward approach is to have them communicate in natural language (which is what many current experiments do, essentially having the agents "chat" with each other in English). This is flexible and leverages LLM strengths, but might be inefficient or prone to misunderstanding without some structure.

This is where an interoperability framework becomes crucial: it can provide a structured medium for agent interaction (e.g., a shared memory or a common set of message types). The \textbf{Coral Protocol} is positioned in this landscape as an infrastructure to help organize and facilitate these multi-agent interactions. It extends ideas from MCP so that agents not only access data/tools uniformly, but also discover and communicate with \textit{each other} in a standardized way. By having a common protocol, an "agent society" can form where each agent knows how to announce its capabilities, listen for requests, and share results in a mutual language. This would allow, for example, an agent built by Company A to collaborate with an agent from Company B if both speak Coral/MCP, much as devices on the internet interoperate via TCP/IP. In summary, multi-agent AI systems are an exciting and fast-evolving area — moving from single, siloed AI assistants towards \textbf{networks of cooperating agents}. The recent progress in this domain (just in the past couple of years) underscores the need for interoperability solutions: to manage complexity, avoid reinventing integration logic, and unlock the full potential of collective AI intelligence. Coral Protocol’s goal of facilitating agent interoperability directly addresses this need, aiming to enable robust, framework-agnostic collaboration among AI agents in the coming "society of AI agents."

\subsection{Agent Communication Languages}
\label{sec:Agent-Communication-Languages} 

In the early development of multi-agent systems, researchers designed specialized agent communication languages to enable structured information exchange between autonomous agents. Two pivotal examples were KQML (Knowledge Query and Manipulation Language) and the FIPA Agent Communication Language (ACL). 
KQML, introduced in the 1990s as part of DARPA’s Knowledge Sharing Effort \cite{Finin1994kqml}, defined a high-level message format and protocol for agents to share knowledge independent of any specific ontology or transport. It centered on an extensible set of message “performatives” (such as ask, tell, achieve, etc.) that represent different types of speech acts an agent can perform. 
Following KQML, the Foundation for Intelligent Physical Agents (FIPA) – an international standards body founded in 1996 – sought to build on these ideas and address some of KQML’s limitations around semantic rigor and standardization \cite{Fipa2002acl}. 
FIPA defined an Agent Communication Language (ACL) that became a widely recognized standard in academic agent research.
FIPA-ACL preserved the use of performatives (e.g. inform, request, confirm, etc.) but provided a more formally defined core ontology and semantics for them. 
This approach, inspired by \textit{speech-act theory}, provided a rigorous way to reason about conversations between agents and was important for research on negotiation, cooperation, and coordination in multi-agent systems. 
The strengths of FIPA-ACL included this formal semantics and a comprehensive framework of interaction protocols (e.g. for contract-net bidding, auctions, query-ref, etc.), which gave developers a blueprint for implementing complex dialogues. 

Despite their conceptual elegance and influence on academic research, neither KQML nor FIPA-ACL achieved widespread adoption in today’s LLM-based agent ecosystems. Limitations became apparent. KQML, while flexible, never fully specified the semantics of its performatives, leading to inconsistencies in how different implementations interpreted messages. FIPA-ACL, on the other hand, was more tightly specified but also more complex: it required developers to commit to a particular set of interaction semantics (the mental-state model) that could be difficult to verify in practice (since one cannot directly inspect another agent’s “beliefs”). 
Both frameworks assumed agents would share common ontologies and trust the communicated mental attitudes, assumptions that are hard to guarantee in open systems. Moreover, the infrastructure envisioned by FIPA (with directory services, agent management systems, etc.) introduced overhead that, outside of research testbeds, proved cumbersome. By the mid-2000s, FIPA as an organization was dissolved without seeing broad industry uptake. 

In the resurgence of AI agents driven by large language models, these older ACLs have not been the foundation — modern agent developers often prefer lightweight JSON or natural-language messaging over the formal, logic-based formats of KQML/FIPA. In summary, KQML and FIPA-ACL were critical steps in multi-agent system development, establishing the idea of structured agent dialogues and common performatives. However, their formality and assumptions (shared semantics, mental state tracking) limited their practicality for the new wave of agents, which operate in more heterogeneous, data-driven environments. This gap has set the stage for a new generation of AI agent communication protocols that seek to combine interoperability with the flexibility required by modern AI systems.

\subsection{Blockchains and Secure Payments}

Blockchains provide a decentralized, tamper‐evident ledger for recording all financial exchanges among agents. By design, every transaction written to the chain is immutable and publicly verifiable, ensuring reliable auditability. 
In Coral Protocol, agent payments (and any inter‐agent transfers) are committed to the blockchain, creating an indelible audit trail of who paid whom and when. This means that all payments for services are transparently recorded and cannot be forged or altered after the fact. 

Beyond simple record‐keeping, smart contracts encode the logic of payments and escrows directly on‐chain. For example, an multi-agent application can publish a task as a smart contract that holds funds in escrow: once the specified work is completed and verified, the contract automatically releases payment to the agents. 
Such contracts allow conditional payments and automated dispute resolution without centralized intermediaries. 
In practice this enables fine‐grained, trustless micropayments – agents can be compensated in real time for individual actions or API calls, and funds are only transferred when task conditions are met. 
These programmable contracts ensure that incentives are aligned: agents are rewarded for correct behavior and cannot receive payment without fulfilling their commitments. 

Together, the blockchain and smart‐contract layers create a fully decentralized trust model for Coral. Users and agents need not trust any single party; instead, they rely on the blockchain’s consensus mechanisms and cryptographic guarantees. 
Every payment is signed and timestamped on chain, so any agent or user can audit the history of transactions on demand. 
In effect, Coral’s \textbf{Secure Payments} service uses the blockchain as a backbone for the agent‐economy: it ensures that all transfers are authorized, transparent, and tamper‑proof. 
This underpins an open marketplace of agent services where developers can list capabilities and set fees, and clients can purchase those services via on‐chain transactions. 
By recording all payments on a public ledger, Coral aligns agents’ incentives (they earn tokens for useful work) and enables flexible, decentralized commerce among agents without relying on traditional financial intermediaries.

\newpage
\section{Motivation}

Integrating AI agents into digital ecosystems isn’t just a technological leap—it’s a glimpse into a future where autonomous systems drive innovation at an unprecedented scale. 
Across the Internet, countless AI agents, often developed by different entities, are already analyzing data, making decisions, and transacting value independently. Yet, most of them operate in isolation, often without mutual trust or a shared understanding.

Now consider the possibilities when these AI agents—each specialized, decentralized, and autonomous—collaborate to break down intricate tasks into manageable components. This showcases the true potential of coordinated AI systems.
Your AI agent could fulfill complex requests by orchestrating a network of agents in real time, negotiating smart contracts across multiple platforms, or collaboratively managing tasks to deliver a unified outcome.
The potential is staggering: a distributed intelligence capable of solving complex, multi-dimensional problems more rapidly and efficiently than any single agent or system alone.

Imagine, for example, spinning up an army of domain-expert AIs as easily as opening an app. A developer commits code and, without lifting a finger, a Git-Diff Reviewer triggers a whole relay: pen-testing, architectural refactoring, unit-test regeneration, performance and accessibility sweeps—the green constellation on the left hums along until every check is green. Across the hall, a hackathon organiser watches a red cluster come alive: an Event Planner talks to an ElizaOS concierge, schedules judges, provisions venues and even fires up an autonomous Event Runner while a friendly UI keeps humans in the loop. Meanwhile the blue B2B-sales galaxy is harvesting leads—in seconds a Deep Researcher, LinkedIn Outreach bot, HubSpot connector and Account-Manager chain track prospects from first touch to closed deal.

Each AI agent acts independently but in collaboration with others, forming an adaptive, intelligent network capable of dynamic, real-time problem solving. 
But such a vision hinges on a critical foundation: \textbf{trust}.

To realize this scenario, we need more than just smart algorithms—we need a \textbf{trustworthy AI agent communication infrastructure}. One where identity is verifiable, intentions are auditable, and interactions are secure and reliable. Only then can we unlock the full potential of multi-agent collaboration on the open Internet and redefine how intelligence is distributed and coordinated at scale.

\subsection{Rising AI Agent Communication Protocols}
\label{sec:Rising-AI-Agent-Communication-Protocols}

As AI agents have re-emerged via large language models and tool-using assistants, a number of modern communication protocols are being proposed to facilitate agent interaction. 
These efforts are motivated by the need to connect agents with data sources, other agents, and services in a standardized way — addressing the shortcomings of both ad-hoc integrations and the older ACLs. 
Three notable emerging protocols are the Agent-to-Agent (A2A) protocol, the Agent Network Protocol (ANP), and AGNTCY’s suite of standards. 
Each takes a distinct approach to agent communication and coordination, and each illustrates both new capabilities and remaining challenges in this space.

\subsubsection{Agent-to-Agent (A2A) Protocol}

Google introduced the Agent-to-Agent (A2A) protocol, an open, vendor-neutral standard designed to enable seamless communication and collaboration among AI agents across various platforms and ecosystems \cite{Surapaneni2025}. 
A2A focuses on task-oriented interactions, providing standardized task objects with clearly defined lifecycles, and supporting the exchange of task-related artifacts to facilitate collaborative task executions. 
Built on established web standards like HTTP, JSON-RPC, and Server-Sent Events (SSE), the protocol ensures compatibility with existing technologies while offering enterprise-grade security through robust authentication and authorization aligned with OpenAPI. Additionally, A2A supports dynamic discovery of agent capabilities via JSON-based "agent cards" and handles both short-term tasks and long-running processes with real-time feedback. Google's collaboration with over 50 industry partners—including Salesforce, SAP, and Workday—highlights the broad industry commitment toward achieving standardized agent interoperability.

In practice, A2A is poised to facilitate complex task executions in both enterprise and open-source contexts. 
In enterprises, multiple AI agents could coordinate processes across business platforms that typically don’t talk to each other today. For example, an agent integrated with a CRM like Salesforce might invoke another agent on an ERP system such as SAP to automatically fulfill an order or update financial records \cite{Gd2025}.
Organizations are exploring scenarios like a sales assistant agent requesting a finance agent to generate a pricing quote, or a customer support chatbot querying an inventory management agent for stock availability – all through A2A’s standardized agent-to-agent calls without human middleware \cite{Justin2025}. 
Because A2A is an open and framework-neutral protocol, it also lends itself to open-source innovation. Developers and researchers can orchestrate modular AI agents (for instance, linking a data analysis agent with a visualization agent) using A2A as the lingua franca for inter-agent dialogue. 
Indeed, popular toolkits like LangChain have begun integrating A2A, allowing independent or specialized agents to be composed into larger AI workflows in research and development settings.

Despite its promise, the A2A protocol has some limitations that motivate the development of more comprehensive multi-agent frameworks. 
Notably, A2A focuses on the mechanics of messaging and doesn’t define a shared ontology or common knowledge base for semantics – agents still must understand the content of messages (often plain text or JSON) based on their own models or agreements, which can limit out-of-the-box interoperability in complex domains. The protocol is also in an early stage (initially released as a draft specification in 2025), so its extensibility to all possible use cases is still evolving and industry adoption is just beginning. 
Moreover, A2A by itself addresses communication rather than higher-level cognition or coordination among agents. It enables agents to talk but does not inherently provide collective planning, sophisticated negotiation, or “hive-mind” reasoning capabilities. 
Implementing such intelligence requires additional layers on top of A2A – for example, orchestration frameworks like Google’s Agent Development Kit (ADK) are intended to manage task executions and decision logic among multiple agents using the A2A channel. 
In summary, A2A lays an important foundation for agent interoperability, but it remains a low-level protocol; achieving robust multi-agent collaboration will also require tackling semantic standards, advanced coordination strategies, and other gaps that A2A alone does not fill.

\subsubsection{Agent Network Protocol (ANP)}

Agent Network Protocol (ANP) aims to become the “HTTP of the agentic web”, explicitly focusing on agent-to-agent communication in a decentralized network. The design of ANP envisions a future where potentially billions of AI agents (from personal assistants to autonomous services) can find each other and exchange messages over the internet just as web services do today. To enable this, ANP provides a multi-layered protocol stack: an identity and authentication layer based on decentralized identifiers (DID) for registering agent identities and establishing end-to-end encrypted channels; a meta-protocol layer that lets agents negotiate which communication protocols or interaction patterns to use with each other dynamically; and an application layer describing the semantics of messages (e.g. agent capabilities, message types, task descriptions) in a standard way. 

In practical terms, ANP’s functionality includes agent discovery (finding agents and publishing one’s availability), agent description (sharing metadata about an agent’s capabilities or APIs, so another agent knows how to interact with it), and a messaging framework that supports various patterns (one-to-one messages, broadcasts, negotiations, etc.) with security and routing handled transparently. An illustrative use case for ANP is a network of cooperative agents in an “Internet of Agents”: for example, a scheduling agent could discover a travel-planning agent and a weather agent, then communicate with both to coordinate a trip itinerary – all through standard ANP messages without custom integration code. Early development of ANP has produced specifications and reference implementations for key pieces, such as an Agent Discovery Protocol and an Agent Description format, and the project has drawn interest in forums like the W3C WebAgents Community Group. The strength of ANP is its ambitious scope: it is explicitly trying to solve inter-agent networking in a general, open way, addressing trust (through decentralized identity and encryption) and scalability (through a common protocol that any platform can implement). By treating agent communication as a first-class internet protocol, ANP is tackling the harder problem of enabling heterogeneous agents to cooperate across organizational boundaries, not just within a single product’s ecosystem. 

Nevertheless, ANP is still in its early stages. Its limitations include the lack of a universally accepted “content language” or ontology for agent messages – while it might standardize the envelopes and routing, agents still need to understand each other’s content (goals, plans, data formats), which is an open challenge. There is also a question of adoption and interoperability: for ANP to succeed as an “industry standard,” it needs many independent agent frameworks to agree on using it. As of now, it remains a promising proposal; it provides pieces of the puzzle (identity, discovery, security), but extensibility and coordination logic are still evolving. Agents using ANP would still need higher-level conventions (negotiation protocols, coordination strategies) defined on top of the base messaging, which are not fully unified. These gaps underscore why new efforts continue to appear – the community recognizes the need for more than just a transport, but a common understanding for agent collaboration.

\subsubsection{AGNTCY (Internet of Agents Initiative)}

Another major initiative gaining traction (launched in late 2024) is AGNTCY – a coalition-driven effort by organizations including Cisco, LangChain, LlamaIndex, Galileo, and others to create an open standard for AI agent interoperability. AGNTCY is often described as laying the foundation for an “Internet of Agents,” drawing a parallel to how early internet standards like TCP/IP and DNS connected disparate systems. The design of AGNTCY actually encompasses multiple complementary protocols and frameworks. At its core are two key specifications: the Open Agent Schema Framework (OASF) and the Agent Connect Protocol (ACP). OASF is essentially a standard schema or metadata format for describing agents – it defines how an agent can publish its capabilities, interfaces, and other characteristics in a machine-readable way. This is crucial for agent discovery and evaluation: if all agents describe themselves using OASF, then a directory or search mechanism can allow agents (or humans) to find suitable agents for a given task and understand how to interact with them. On the other hand, ACP is the communication protocol that allows agents to invoke and interact with each other once they’ve discovered each other. It covers establishing connections, authentication/authorization handshakes, exchanging messages or task instructions, and handling errors in a standardized way – effectively, ACP aims to let an agent built on one framework “call” an agent built on another, as seamlessly as a function call or API request. 

Together, these pieces enable what AGNTCY envisions: for example, an organization could deploy multiple AI agents with different specialties (say, an HR assistant, a coder agent, and a data analyst agent) and, using AGNTCY standards, have them discover each other, share tasks, and coordinate task executions even if they were built by different vendors. 
Some early demonstrations (and code releases in 2025) show agents registering in a common directory and then composing their abilities to solve multi-step problems, illustrating the potential of this interoperability. The design goals of AGNTCY heavily emphasize extensibility and community governance. Rather than dictating a single closed protocol, the initiative encourages developers to extend the specs and contribute new ideas, hoping to avoid fragmentation by uniting efforts under one open umbrella. Its backers liken its importance to that of fundamental internet protocols, arguing that a similarly neutral and extensible standard for AI agents will unlock innovation across industries. 

As of early 2025, AGNTCY is still ramping up; its initial specifications are available and open-source, but it acknowledges that broad adoption is crucial for success. 
In terms of limitations, AGNTCY’s challenges echo those of any nascent standard: it currently lacks widespread implementation, and there is a risk of competing protocols (like those above) dividing the community. 
The effort needs to onboard many AI platforms and toolmakers to truly become the “standard” – in other words, it faces an uphill battle for critical mass, where if not enough parties adopt it, the goal of universal agent interoperability could fail due to multiple incompatible ecosystems. 
Additionally, while AGNTCY covers discovery and connection, it is still developing the full “coordination logic” for complex task exeuctions (their roadmap includes stages for orchestration and monitoring of multi-agent systems). 
This means that questions of how agents plan joint tasks, agree on protocols for negotiation, or maintain long-term collaborations may require further conventions on top of AGNTCY’s base layer.

\subsubsection{NANDA (Networked Agents and Decentralized AI)}

NANDA is an MIT Media Lab initiative that envisions an “Internet of AI Agents” – a global, decentralized network where agents can discover, communicate, and transact autonomously. 
Architecturally, NANDA is described as a “rules-based operating system for agents”. 
It employs a multi-layered protocol stack built on existing standards (e.g. Anthropic's Model Context Protocol) and adds comprehensive identity, discovery, and trust mechanisms. For example, agents register themselves in decentralized registries and are authenticated by cryptographic certificates. 
NANDA integrates identity management, verifiability, and portable reputation within the protocol so that any agent’s identity and track record can be verified by others. 
In this way, agents can self-register and discover each other’s capabilities without a central authority, and then establish secure, end-to-end communication channels for coordination.

NANDA's core goals include secure multi-agent coordination and composability. 
It provides standard data schemas and message formats so that agents from different domains can understand one another. 
Agents describe their capabilities via a common schema framework, much like a machine-readable profile, and then use NANDA's protocols to invoke and integrate those capabilities. 
For instance, an agent can query a registry to find specialist agents, then invoke them in sequence to complete a complex task. Security and accountability are built in: interactions and agreements can be enforced via cryptographic proofs, and an agent’s portable reputation token accrues across transactions. 
In effect, NANDA turns every API or data service into an interactive network participant with verifiable identity, enabling agents to compose arbitrary services into workflows in a trust-minimized environment.

Compared with Coral, NANDA takes a more “full-stack” approach to interoperability. Both systems aim to let heterogeneous agents work together, but they emphasize different layers. 
Coral Protocol relies on the Open Agent Schema Framework (OASF) and the Agent Connect Protocol (ACP) to define a uniform way for agents to publish capabilities and call one another’s interfaces. 
In Coral, any agent that implements the standard schemas can be invoked like an API, and payments for services are handled via the blockchain. 
NANDA, by contrast, embeds additional layers into the fabric of the network: it assumes built-in decentralized discovery, identity and reputation layers, and even governance protocols. 
In other words, NANDA seeks to standardize the entire agent ecosystem (from messaging to economics) by consensus-driven rules, whereas Coral focuses on modular protocols and a token-backed marketplace to enable interoperability. 
Both envision a decentralized agent economy, but NANDA leans more on architectural standardization (certificates, registries, consensus), while Coral leverages blockchain payments and open, composable interfaces to achieve similar goals.

\subsubsection{Synergetics.ai for AI Agents}

Synergetics.ai\footnote{\url{https://synergetics.ai}, accessed on April 28, 2025.} is a startup focused on secure, decentralized communication and commerce for AI agents. 
In early 2025 it introduced AgentWorks™, a suite of tools designed to let agents operate across enterprise boundaries. The suite includes services for identity (AgentID), discovery (Agent Registry), connectivity (AgentConnect), and digital wallets, but its core is the AgentTalk protocol – a patented, extensible agent-to-agent messaging and transaction layer. Synergetics envisions giving every agent a verifiable identity (for example using a “.TWIN” blockchain domain as a wallet) so that agents can authenticate and transact without human intervention.

The design goals of Synergetics emphasize security, interoperability, and decentralized trust. Agents created on the platform are provisioned with cryptographic IDs (AgentID) and zero-knowledge proof credentials to enable Know-Your-Agent (KYA) verification. Communications over AgentTalk are end-to-end encrypted, and transactions between agents (such as service purchases or data exchanges) can be anchored on blockchain using smart contracts. In practice, Synergetics provides a layered stack: for example, AgentRegistry handles discovery and trust, AgentWallet manages keys and tokens, and AgentConnect bridges agents to web and metaverse environments. According to Synergetics, its AgentTalk protocol (also referred to in their architecture discussions as “AgentFlow”) handles real-time, asynchronous inter-agent messaging and workflow coordination. This approach is akin to an OSI-inspired model: lower layers ensure transport and identity, while the AgentTalk/AgentFlow layer provides a common “language” and control structure for agents to negotiate tasks.

Synergetics’s platform exhibits several notable strengths. By integrating identity, registry, communication, and even a marketplace (AgentMarket) under one framework, it ensures full traceability and security of agent interactions. Every agent effectively has a “passport” (the .TWIN DNS name) and a built-in digital wallet, which enables secure transactions and ownership of AI assets. The use of blockchain and tokenization is innovative: for example, Synergetics touts that agents can be listed, bought, or subscribed to on an open marketplace, with royalties and subscriptions enforced on-chain. The platform explicitly targets interoperability (even with embedded and physical agents) and real-time automation, as emphasized in its marketing materials. In short, Synergetics offers a practical, end-to-end agentic ecosystem where agents can find each other, negotiate terms, and carry out complex workflows across organizations.

However, Synergetics also has limitations. It is a single-vendor, proprietary system (with key technologies patented), so broad community-driven standardization is lacking. Adoption to date is still nascent, and integrating its blockchain-based identity stack could present performance or complexity challenges for some users. Like other early proposals, Synergetics prioritizes transactional and identity layers; it does not itself define high-level semantics or reasoning ontologies for agents, leaving rich multi-agent reasoning to be built on top. In practice, its focus on market-driven agent capabilities (agents as products) means it may be less suited for purely academic or open-source scenarios than consortium-led efforts.

Compared to the other approaches discussed above, Synergetics overlaps with them in some aspects but diverges in others. For example, Google’s A2A also standardizes messaging but lacks a built‑in identity or economic layer – whereas Synergetics bakes both into its fabric. Like ANP and NANDA, it employs decentralized identifiers and registries for agent discovery, but Synergetics ties them directly to blockchain DNS and wallets. Unlike the multi-organization AGNTCY initiative, Synergetics delivers a complete operational stack (with wallets and a marketplace) rather than just schemas and connectors. In summary, Synergetics demonstrates a commercially viable vision of an “agent economy,” yet remains a closed ecosystem. Its innovations in security and agent commerce are complementary to, but not a replacement for, open protocols. These features and trade-offs will inform the design of Coral Protocol, which in the next section we propose as a more unified, extensible agent communication layer.

\subsubsection{IBM watsonx Orchestrator}

IBM introduced \textit{watsonx Orchestrator}\footnote{Also marketed as \textit{IBM watsonx Orchestrate}.} as an enterprise-focused platform for multi-agent workflow automation and coordination \cite{ibm2025watsonx}.  Rather than defining just a communication protocol, watsonx Orchestrator delivers an end-to-end environment to build, run and manage AI agents that can collaborate across diverse business applications and data sources. Key capabilities announced in 2025 include a no-code Agent Builder (enabling custom AI agents to be created in under five minutes) and an Agent Catalog listing over \textbf{150} pre-built agents and tools from IBM and partners \cite{ibm2025watsonx}.  IBM provides domain-specialized agents (e.g. for HR, procurement, sales) with ready-made skills, alongside utility agents for tasks like web research and calculations, to jump-start common enterprise use cases \cite{ibm2025watsonx}.  These agents—and any custom or third-party agents—are orchestrated through a unified interface, with a central AI-driven planner intelligently routing tasks to the appropriate agent, tool, or human stakeholder as needed. Powered by large language models (e.g. IBM’s Granite series) for reasoning and planning, the Orchestrator acts as a “multi-agent supervisor, router and planner,” autonomously coordinating agents to complete complex workflows while abstracting away the complexity for end users \cite{ibm2025watsonx}.

A core strength of watsonx Orchestrator is its deep integration into enterprise IT ecosystems.  The platform comes pre-integrated with 80+ enterprise applications and services (Salesforce, SAP, Workday, Microsoft tools, etc.), allowing AI agents to directly interact with existing software stacks for actions like database updates, form submissions, or ticket handling \cite{ibm2025watsonx}.  This out-of-the-box connectivity, combined with support for hybrid cloud and on-premise deployments, reflects IBM’s emphasis on immediate business value—agents can “plug into” legacy systems and cloud services without extensive custom integration. In contrast to pure protocol efforts, IBM’s solution provides built-in workflow orchestration: it not only enables messaging between agents but can automatically sequence multi-step processes (planning, invoking one agent after another, aggregating results, handling exceptions) to fulfill a high-level task. Enterprise-grade features such as agent observability and governance are also included \cite{ibm2025watsonx}. For example, watsonx Orchestrator supplies monitoring dashboards and guardrails to track agent performance, data usage, and compliance, leveraging IBM’s AI governance frameworks for risk management. This focus on management and oversight addresses concerns like reliability, security and accountability which are critical in corporate settings.

Although IBM’s offering is a proprietary platform, it is designed with interoperability in mind. The Orchestrator’s extensible architecture allows agents built on various frameworks (including open-source agent libraries like LangChain, LangGraph, etc.) to be onboarded via standardized interfaces. IBM has introduced an \textit{Agent Connect} developer framework that exposes familiar REST/JSON and chat-based APIs for external agents to communicate within the Orchestrator environment. Notably, watsonx Orchestrator also embraces emerging agent communication standards: it supports Anthropic’s Model Context Protocol (MCP) for tool and data access interoperability, and IBM has open-sourced an \textit{Agent Collaboration Protocol (ACP)} to facilitate agent-to-agent messaging within its ecosystem \cite{ibm2025watsonx}. By aligning with standards like MCP (and potentially bridging to protocols like A2A), IBM aims to avoid isolating its agents from the wider “internet of agents.” In essence, the platform implements a superset of agent communication functionality — a practical application layer that sits atop low-level protocols, abstracting them into a cohesive workflow engine.

\subsection{Why Coral Protocol?}
The AI‑agent ecosystem is rapidly diversifying. Google’s Agent‑to‑Agent (A2A) protocol and the open‑source Agent Network Protocol (ANP) introduce common schemas and decentralized identity layers so agents can ``speak’’ a shared language. \textbf{IBM’s watsonx Orchestrator} demonstrates an enterprise‑grade, no‑code environment that plans and supervises multi‑agent workflows while offering deep, out‑of‑the‑box integrations with systems like Salesforce, SAP, and Workday. Meanwhile, consortium efforts such as Cisco’s AGNTCY and MIT’s NANDA outline ambitious blueprints for a decentralized \emph{Internet of Agents}, and commercial stacks like Synergetics.ai add wallet‑centric marketplaces.  

Yet each of these initiatives addresses only a slice of the full problem space.  
\begin{itemize}
  \item \emph{A2A} excels at framework‑agnostic message exchange but is point‑to‑point and leaves incentives and team logic to higher layers.  
  \item \emph{ANP} provides strong DID‑based identity and peer discovery, yet omits built‑in payment rails or economic coordination.  
  \item \emph{IBM watsonx Orchestrator} offers AI‑driven planning and governance dashboards, but—as a proprietary platform—cannot guarantee vendor‑neutral participation or tokenised incentives beyond IBM’s ecosystem.  
  \item \emph{AGNTCY} and \emph{NANDA} articulate open registries and governance but remain largely conceptual, with limited turnkey tooling for secure team assembly or on‑chain value transfer.  
\end{itemize}

\noindent\textbf{Coral Protocol} closes these gaps by combining the strongest ideas from prior work into a cohesive, open stack:

\begin{itemize}
  \item \textbf{Vendor‑neutral interoperability and extensibility}.  
        Coral re‑uses A2A‐style capability cards and ANP‑inspired DID schemas, allowing any agent framework (including watsonx agents) to plug in without rewriting core logic.  

  \item \textbf{Secure team formation}.  
        Agents cryptographically bind into ad‑hoc coalitions using multi‑party signatures and shared DIDs, enabling large‑scale task forces that no point solution currently supports.  

  \item \textbf{Integrated payments and incentives}.  
        Every agent or tool can publish services in a shared market and receive on‑chain micro‑payments, turning collaboration into an economy—something absent from A2A, ANP, and watsonx Orchestrator alike.  

  \item \textbf{Dynamic multi‑agent teams}.
        Discovery, delegation, and result aggregation are baked into the runtime so that complex objectives can be decomposed across many agents—extending watsonx‑style orchestration beyond a single vendor’s boundary.  

  \item \textbf{Open, decentralized governance}.  
        Specifications, reference code, and registries are stewarded in a neutral foundation, ensuring that no single corporation controls the evolving ``Internet of Agents’’ fabric.  
\end{itemize}

\noindent In short, Coral Protocol is not merely another niche solution—it is designed as the unifying substrate that weaves together messaging, security, economic alignment, and open standards, enabling heterogeneous AI agents and humans to cooperate at global scale.

\newpage
\section{The Coral Ecosystem}

\begin{figure}[ht!]
    \centering
    \includegraphics[width=1.0\linewidth]{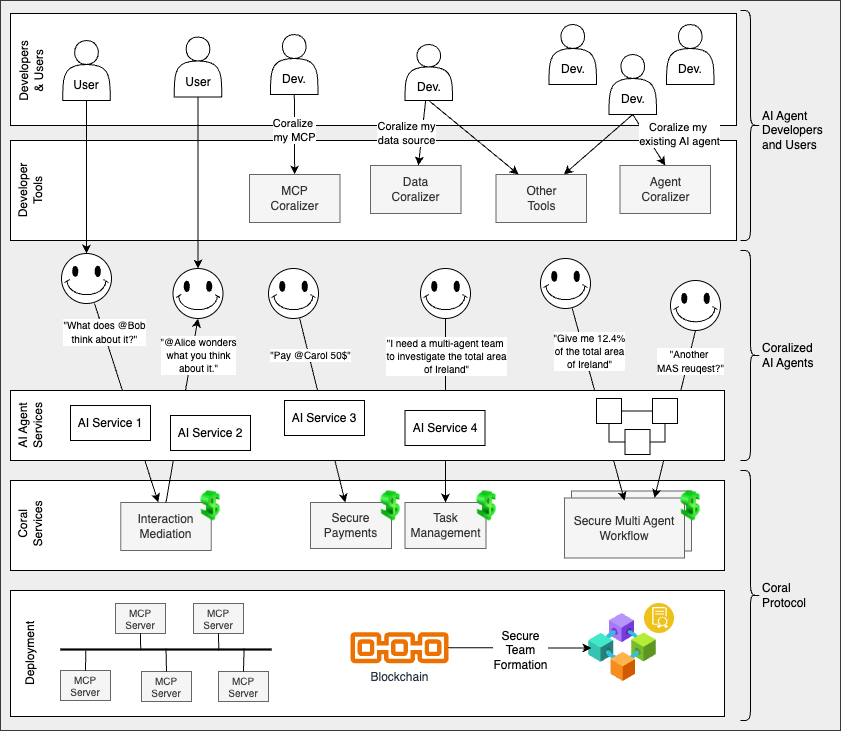}
    \caption{The Coral Ecosystem}
    \label{fig:Coral-High-Level-Architecture}
\end{figure}

The Coral ecosystem is organized into three major contexts: \textit{AI Agent Developers and Users}, \textit{Coralized AI Agents}, and \textit{Coral Protocol} (see Figure \ref{fig:Coral-High-Level-Architecture}). These contexts encapsulate the roles, components, and interactions that enable Coral's secure and interoperable multi-agent environment. In this section, we describe each context and its components in detail, and explain the data flow and interactions between them in the overall architecture.

\subsection{AI Agent Developers and Users}
\label{sec:AI-Agent-Developers-and-Users}

This context represents the human actors (developers and end-users) and their toolchain for interfacing with the Coral ecosystem. Developers contribute to Coral by integrating external AI capabilities and resources. They use specialized \emph{Coraliser} modules to onboard different assets into the ecosystem: the \textbf{MCP Coraliser} connects external model endpoints via the Model Context Protocol (MCP), allowing external AI models or services to communicate in Coral’s standard format; the \textbf{Data Coraliser} links external data sources (e.g., databases, knowledge bases or live data streams) into the Coral ecosystem, making those data accessible to AI agents; and the \textbf{Agent Coraliser} wraps existing AI agents or services (such as pre-existing language models or automation scripts) to comply with Coral’s protocols and interfaces. Through these coralisers, developers effectively “Coralise” their models, data, or agents — registering them into the ecosystem so that they become available as Coralized AI Agents in the middle context. In addition to these, developers can utilize \textbf{Other Tools} (e.g., software development kits, testing frameworks, or monitoring dashboards provided by the Coral platform) to build, debug, and optimize their agents and integrations. 

\textbf{End-users} in this context are the consumers of the AI agent services. They interact with Coralized AI Agents by issuing natural language queries or commands through user-facing applications or interfaces built on top of Coral Protocol. Users need not be aware of the underlying complexity; they simply pose questions or tasks in everyday language. For instance, a user might ask, “What does AgentX think about hypothesisY?” or instruct the system with a command like, “Pay \texttt{@AgentZ} \$50 to investigate the total area of region~R.” These inputs from users enter the Coral ecosystem via the Interaction Mediation service of Coral Protocol (described later), which ensures the queries are routed appropriately. In summary, the Developers and Users context supplies the ecosystem with integrated AI capabilities (via coralizers) and the driving queries or tasks (from users) that initiate agent interactions.

\subsection{Coralised AI Agents}
\label{sec:Coralized-AI-Agents}

The middle tier of the architecture consists of the Coralised AI Agents — the ensemble of AI services and agents that have been onboarded into the ecosystem and operate under Coral’s framework. Each Coralised agent is an AI service (often backed by a large language model or specialized AI module) that adheres to Coral Protocol for communication and security. Once developers coralise a model, data source, or existing agent, it becomes a Coralised AI Agent accessible in this layer. Collectively, these agents form a distributed, interoperable multi-agent system, each capable of understanding and responding to natural language prompts.

When an end-user query or command enters the system (for example, the questions or instructions mentioned above), Coral Protocol’s Interaction Mediation component will dispatch it to the appropriate agent or agents in this layer. An agent specialized in a certain domain (say, Agent~X in the earlier query) will receive the question directed at it and generate a response drawing on its knowledge or data (which might have been linked via a Data Coralizer). 
Coralized agents can also initiate interactions with one another under guidance of Coral Protocol. 
For instance, if a user’s request is complex and requires multiple skills, Coral Protocol can intelligently form a dedicated team by discovering and composing the necessary AI agent together. All such inter-agent dialogues are mediated and logged by the protocol to maintain coherence and security.

Crucially, Coralized AI Agents can handle not only informational queries but also procedural commands that involve actions in the ecosystem. For example, when a user says “Pay \texttt{@AgentZ} \$50 to perform task T,” the addressed agent (Agent~Z) will be invoked to execute the task T, and Coral Protocol will engage its secure payment service to transfer the specified amount. Throughout their operation, coralized agents remain decoupled from any particular user interface or platform — they rely on Coral Protocol to handle incoming requests, outgoing replies, and any coordination with other services. This design allows a heterogeneous collection of AI agents to work together seamlessly: each agent focuses on its specialized processing (e.g., analyzing data, answering questions, performing computations), while the protocol manages communications and shared context. The result is a flexible multi-agent ecosystem where, from the user’s perspective, complex tasks (potentially involving several agents and data sources) can be invoked with simple natural language requests.

\subsection{Coral Protocol}

Coral Protocol is the underlying framework that connects users, developers, and coralized agents, providing the services and infrastructure necessary for secure, coordinated interactions. It comprises two main parts: a set of core \textit{Coral Services} that mediate and secure the ecosystem’s operations, and the \textit{Deployment} and infrastructure layer that supports these services. An important additional concept in this context is \textbf{Secure Team Formation}, which refers to the protocol’s capability to dynamically assemble multiple agents into collaborative task teams in a secure manner. We describe each of these aspects below.

\subsubsection{Coral Services} 
\label{sec:Coral-Services} 

Coral Protocol offers several key services to orchestrate interactions: 

\begin{enumerate}
\item \textbf{Interaction Mediation} is responsible for routing and managing all messages between users and agents (and between agents themselves). It receives user queries or commands from the interface and determines which Coralized AI agent(s) should handle them, forwarding the query along with any relevant context. It also ensures that responses or follow-up questions from agents are delivered to the right recipients (e.g. back to the user or to another agent), maintaining the dialogue state. 
\item \textbf{Secure Payments} handles monetary transactions and incentives within the ecosystem. When a user’s instruction includes a payment (such as paying an agent for a service), this service securely executes the transaction, escrow if necessary, and confirms the payment using the underlying blockchain. It ensures that financial exchanges between users and agents (or between agents) are authorized and recorded, enabling a marketplace of agent services. 
\item \textbf{Task Management} oversees the life-cycle of complex tasks that agents undertake. It assigns or schedules the sub-tasks to the appropriate agents, monitors their progress, and aggregates results. For example, if a user’s query spawns a multi-step research task involving several agents, the Task Management service will track each step and ensure the overall task completes successfully. 
\item \textbf{Secure Multi-Agent Teamwork} coordinates scenarios where multiple AI agents collaborate on a shared objective. This service sets up and manages the team of information and control between agents, enforcing security policies such as authentication of each agent’s identity and permissions. It ensures that agents exchange data through approved channels and that the execution follows any constraints (for instance, not revealing sensitive data to an agent lacking clearance). Together, these Coral services form the intelligent control plane of the ecosystem, mediating every interaction to guarantee it is executed efficiently and safely.
\end{enumerate}

\subsubsection{Deployment Infrastructure} 
\label{sec:Deployment-Infrastructure} 

To support the above services and the execution of AI agents, Coral Protocol leverages a distributed deployment infrastructure. A network of \textbf{MCP Servers} hosts the AI models and agent runtimes. “MCP” refers to the Model Context Protocol – a standardized interface that these servers implement to allow agents to request model inference or tool usage with a unified API. By deploying agents on MCP servers, Coral ensures that each agent can be invoked remotely with proper context and resource allocation, abstracting away the hardware or environment details. These servers handle the heavy computation of AI tasks (such as running large language model inferences or data processing jobs) and expose endpoints for the Interaction Mediation service to call. In parallel, the Coral ecosystem integrates a \textbf{blockchain} network into its infrastructure. The blockchain serves as a secure ledger for recording important events and transactions: it records payment transactions issued via the Secure Payments service, and can also log multi-agent agreements or task completion records for auditability. By using blockchain technology, Coral introduces an immutable and transparent layer of trust – for example, users and developers can verify that an agent was indeed paid for its service, or that a particular multi-agent task’s outcome was agreed upon by all parties. The blockchain also aids in decentralized governance of the ecosystem, ensuring no single party can tamper with the history of agent interactions or outcomes.

\subsubsection{Secure Team Formation} 

An advanced feature of Coral Protocol is its support for secure team formation, which is the dynamic assembly of multiple agents into a collaborative team to solve complex tasks. 
When Interaction Mediation and Task Management determine that a user’s request requires diverse expertise (for instance, a task needs one agent to gather data, another to analyze it, and another to verify the results), the protocol initiates secure team formation. 
This process involves selecting the appropriate set of Coralized AI Agents, establishing authenticated communication channels among them, and defining each agent’s role in the complex task. 
The \emph{Secure Multi-Agent Team} and blockchain infrastructure play pivotal roles here: each agent’s identity and permissions can be verified against blockchain records or certificates, and any inter-agent contracts (such as payment splits or data access agreements) can be codified as smart contracts or logged transactions. 
The result is a trusted ad-hoc team of AI agents that can cooperatively work on the user’s task. Throughout the team’s operation, the Coral services enforce that information sharing is restricted to what is necessary for the task and that any interim results are reported back to the Task Management module. 
Once the collaborative task is complete, the team can be disbanded, with the outcome delivered to the user as a coherent result. 
Secure team formation thus enables scalability of problem-solving: the ecosystem can marshal multiple specialized agents together, while maintaining security and trust among agents that may have been contributed by different developers or organizations.

\subsection{Inter-Context Data Flow and Interaction} 
\label{sec:Inter-Context-Data-Flow-and-Interaction} 

Bringing together the above elements, the data flow in the Coral ecosystem moves fluidly across the three contexts under the governance of Coral Protocol. A typical interaction begins with an end-user query or command from the Developers/Users context. 
This request enters Coral Protocol, which uses its Interaction Mediation service to interpret and route the request to one or more Coralized AI Agents. Those agents, running on MCP Server infrastructure, receive the query along with any additional context or data provided (possibly accessing external data through integrated sources via Data Coralizers). 

As the agents process the query, they may perform computations, look up information, or even generate sub-queries to other agents. 
Any such agent-to-agent interactions are again managed by the protocol's services. 
If the user's request entails a transactional component (e.g., paying an agent or purchasing data), the Secure Payments service engages with the blockchain to execute and record the transaction, ensuring the agent proceeds with the task once payment is confirmed. 
For complex requests, the Task Management service may split the work among multiple agents and invoke secure team formation to coordinate them. Intermediate results flow back through the Task Management and Interaction Mediation layers, which assemble the final answer or result. 

Finally, the response is delivered back to the end-user, completing the round-trip. Throughout this process, each component in the Developers/Users context (such as a developer’s integrated agent or data source) and each Coralized agent operates within the rules of Coral Protocol, ensuring interoperability. The high-level architecture diagram (Figure~\ref{fig:Coral-High-Level-Architecture}) encapsulates these interactions, showing how the human-facing side, the AI agent side, and the protocol services side all connect. In essence, the Coral ecosystem’s architecture enables a seamless and secure loop: humans provide instructions and integrations, AI agents provide intelligence and action, and Coral Protocol ties everything together with standardized communication, security, and coordination mechanisms suitable for a robust multi-agent system.

\newpage
\section{Coral Protocol Architecture}

\subsection{Overview}

Coral Protocol’s architecture is organized into layers that connect end-user applications, developer tools, AI agents, and shared infrastructure. 
At a high level, user applications and developer tooling (top layer) interact with Coral Servers on each host, which in turn coordinate Coralized Agents (middle layer) running on distributed compute servers. 
These components communicate over the Internet and record trust-critical events on a common blockchain (bottom layer). Together, this architecture ensures that agents “wrapped” for Coral (the Coralized Agents) can be invoked and orchestrated in a uniform, secure manner.

\begin{figure}[ht!]
    \centering
    \includegraphics[width=1.0\linewidth]{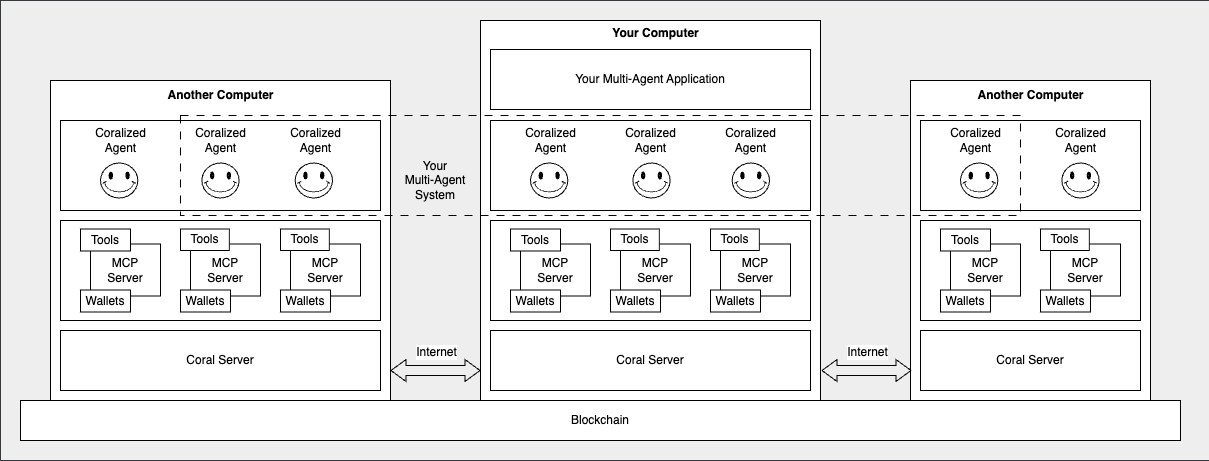}
    \caption{Coral Protocol Architecture}
    \label{fig:Coral-Architecture}
\end{figure}

The architecture diagram of Coral Protocol is shown in Figure \ref{fig:Coral-Architecture}. 
There are multiple computers (nodes) running Coral Servers, each hosting Coralized Agents, MCP servers, development tools, and wallets, all connected via the Internet and secured by an underlying blockchain.

\subsection{Coralized Agents}

The Coralised Agents reside in the middle tier of the architecture and are the core AI services of the system. 
Each Coralised Agent is an AI module or service (often backed by a large language model or specialized analytics tool) that has been onboarded to comply with Coral's protocols (see Section \ref{sec:Coralized-AI-Agents}). 
Once a model or data source is coralised, it becomes a first-class agent in the ecosystem. 
Collectively, these agents form a distributed, interoperable multi-agent system that can understand natural-language queries and perform tasks. When a user request or command arrives (via a user’s application), the Coral Server’s Interaction Mediation service dispatches the query to the appropriate Coralised Agent(s). 
For example, a query directed at “Agent X” is routed to that agent’s instance, which then processes the query and generates a response using its internal knowledge or tools. 
Coralised Agents can also communicate with one another (e.g., to handle sub-tasks in a complex task execution) under the coordination of Coral Protocol's multi-agent teamwork services. In summary, the Coralised Agents layer encapsulates all AI services in the ecosystem, enabling them to interoperate seamlessly (see Section \ref{sec:Coralized-AI-Agents}).

\subsection{MCP Servers and Tools}

The MCP Servers and Tools layer provides the compute and integration endpoints that Coralized Agents use to perform tasks. 
Coral leverages the Model Context Protocol (MCP) as a unified interface (see Section \ref{sec:Model-Context-Protocol-(MCP)}) to link agents with models and external tools. 
In practice, each agent is deployed as a service on one or more MCP servers. 
A network of MCP servers hosts AI model runtimes and heavy computation back-ends, exposing them through a standardized API. 
For instance, an MCP server might run a large language model or a data-processing pipeline; Coral Protocol can invoke it remotely via the MCP interface. 
Notably, MCP supports not just model inference but also tool usage: agents can request to use external tools (APIs, databases, web search, etc.) through the same MCP API. 
In effect, “tools” in the diagram refers to additional functionalities or APIs that agents can call. 
By abstracting hardware and environment details, MCP servers ensure that any Coralized Agent can run anywhere in the network with proper context and resources. 
In short, the MCP Servers and Tools layer supplies the underlying AI models and function endpoints that agents use to generate intelligent results.

Each node also includes a Wallet component to support secure economic transactions. Coral’s built-in Secure Payments service (Section \ref{sec:Coral-Services}) enables users and agents to exchange payments for services, and this requires wallet accounts. 
A wallet holds the cryptographic keys and token balance for a user or agent. 
When an agent completes a paid task, the Secure Payments service uses the wallet to sign and broadcast a blockchain transaction; similarly, users’ wallets are debited when they request a paid service. By tying payments to wallets, the protocol ensures that every financial exchange is authorized and recorded. 
In the architecture diagram, the “Wallets” box under each agent/MCP server indicates that each agent (and user interface) has an associated wallet address. These wallets interface directly with the Coral Server’s payment logic and the blockchain layer, so that statements like “Pay @AgentZ \$50” in a user’s query can be carried out via smart contract or transaction.

\subsection{Coral Server}

The Coral Server is the local host process on each computer that implements the core Coral protocol services (Section \ref{sec:Coral-Services}). 
In the diagram, each node’s large block labeled “Coral Server” contains these services. 
The Coral Server listens for incoming requests from user apps or other agents and uses the Interaction Mediation service to manage message \textit{threads}. 
For example, in a conversation thread, it creates threads, adds participants (agents), and delivers messages (as seen in example in Section \ref{sec:Exploiting-Coral-Protocol-to-a-Real-World-Application}). 

Internally, the Coral Server also invokes the Task Management service to decompose complex jobs into sub-tasks, and the Secure Multi-Agent Teamwork to coordinate inter-agent protocols. 
If a transaction is involved, the Coral Server triggers the Secure Payments service to engage with the blockchain. 
In effect, the Coral Server is the control-plane nexus that ties user queries to agent actions: it routes inputs to the correct Coralised Agents, aggregates their outputs, and enforces security and workflow policies as described in Section \ref{sec:Coral-Services}.

\subsection{Multi-Agent Application}

The Multi-Agent Application layer sits atop Coral Protocol on the developer's computer. 
This is the developer’s or end-user’s application that uses Coral’s APIs to create and orchestrate agents. 
Typically, the application will instantiate one or more “UI agents” or orchestrator agents that interface with the human. 
For example, a UI agent can receive a user’s command (“Compute X for me”) and call the Coral Server to create a new task thread mentioning the required agents. 

As discussed in Section \ref{sec:AI-Agent-Developers-and-Users}, developers use Coralizer modules and SDKs (under Tools) to integrate their agents into this application layer. 
Once a conversation thread is established, the application will relay the user’s natural-language query into Coral’s Interaction Mediation, and later collect the agents’ replies to present back to the user. 
In short, the multi-agent application is the entry and exit point on the user side: it supplies queries into the ecosystem (via the Coral Server) and receives formatted answers from the agents.

\subsection{Internet Communication Layer}

The Internet Communication Layer (indicated by the arrows labeled “Internet” in the diagram) enables distributed operation. 
Coral is inherently network-agnostic: Coral Servers, agents, and MCP servers on different machines communicate over standard Internet protocols (e.g. HTTP or WebSockets). 

In practice, a Coralized Agent running on one host can be invoked by a Coral Server on another machine, as long as they are connected via the Internet. 
The Interaction Mediation service abstracts these network links, so that sending a message to an agent transparently traverses the Internet if needed. 
This layer ensures that no matter where developers deploy their agents or servers, the Coral architecture remains coherent. 
As summarized in Section \ref{sec:Inter-Context-Data-Flow-and-Interaction}, data flows “fluidly across the three contexts” over these network connections.

\subsection{Blockchain Layer}

Finally, the Blockchain Layer underpins the entire ecosystem. Coral integrates a blockchain network into its deployment fabric (see Section \ref{sec:Deployment-Infrastructure}) to provide an immutable ledger. 
All important events—such as secure payment transfers, agent agreements, or execution checkpoints—can be logged on the blockchain. 
This means, for example, that every completed transaction or multi-agent contract is recorded transparently: users and developers can later verify that a payment was made or a task was approved by all parties. 
Because the blockchain is decentralized, it guarantees that no single Coral participant can alter these records. 

In Figure \ref{fig:Coral-Architecture}, the blockchain is shown as the foundation. 
In operation, whenever the Coral Server executes the Secure Payments service, it generates a signed transaction that is broadcast to this blockchain. 
The blockchain also supports the Secure Team Formation process by storing agent identities, permissions, or smart-contract bindings, thereby enforcing trust across different organizational domains. 

In sum, the blockchain layer provides the trust and audit backbone for all Coral transactions and multi-agent tasks executions.

\newpage
\section{Coralised Agents and Coralisation}
\label{sec:Coral-Agents-and-Coralisers}

This section introduces the key components that enable integration and interaction in the Coral ecosystem: \textbf{Coralised Agents}, the active entities that perform tasks and collaborate, and \textbf{Coralisation}, the process that onboard external AI agents and tools (i.e. MCP servers) into the protocol.

\subsection{Coralised Agent}

\begin{figure}[ht!]
    \centering
    \includegraphics[width=0.8\linewidth]{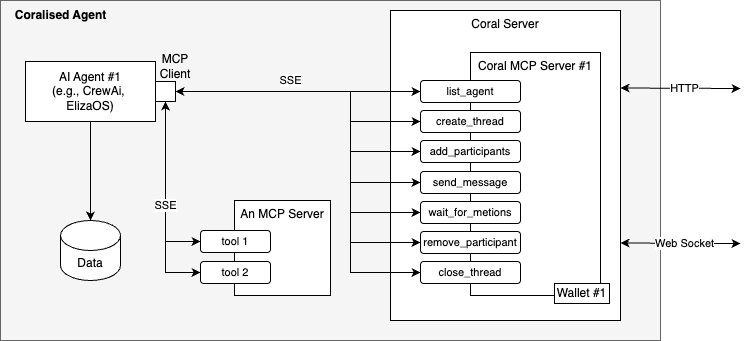}
    \caption{Coralised Agent Architecture}
    \label{fig:Coral-Agent}
\end{figure}

A \textit{Coralised Agent} is an AI agent integrated into Coral Protocol framework, combining an autonomous reasoning engine with communication, coordination, and trust infrastructure. 
Figure~\ref{fig:Coral-Agent} illustrates the architecture of such an agent, which is composed of several modular components: the AI agent itself (e.g., built using frameworks like CrewAI\footnote{CrewAI, \url{https://www.crewai.com}, accessed on June 11th, 2025.} or ElizaOS\footnote{ElizaOS, \url{https://www.elizaos.ai}, access on June 11th, 2025.}), an internal data storage module (optional), MCP server(s) providing tool access, and the Coral server backend providing a Coral MCP server and a dedicated cryptographic wallet for each AI agent.

These components work in unison to enable the AI agent to collaborate with other Ai agents in a thread-based messaging environment. 
All interactions are governed by well-defined protocols and interfaces, ensuring that agents remain loosely coupled yet interoperable within the larger multi-agent system. In particular, communication is facilitated through standard web mechanisms (HTTP, WebSocket) and event streams (Server-Sent Events), which provide a robust, asynchronous messaging substrate. The following provides a detailed description of each architectural element and their interplay in the Coralised Agent framework.

\subsubsection{AI Agent and Data Storage:} 

At the core of the architecture is the AI agent, the autonomous decision-making entity. This agent can be implemented using any suitable agent framework or language model platform (for example, CrewAI or ElizaOS), and it embodies the logic for interpreting inputs, maintaining conversational context, and generating outputs. The agent is designed to operate independently, encapsulating its own goals and cognitive processes. To support long-term reasoning and context management, the AI agent can be paired with an internal \textit{data storage} component. This data store (often a database or vector memory) can hold the agent's knowledge base, conversation history, and any persistent state or embeddings needed for the agent's tasks. By persisting relevant information, the agent can retrieve facts, past dialogues, or domain-specific data as needed, thereby maintaining coherence over extended interactions. The AI agent can query and update this storage as it reasons, effectively using it as its memory. 
The agent does not directly invoke other agents' internals; instead, it interfaces with the outside world exclusively through the standardized channels provided by the Coral framework, which preserves modularity and independence across agents.

\subsubsection{Coral Server (Coral MCP Server)}

At the center of the Coralised Agent framework is the \textit{Coral server}, a dedicated server that enables multi-agent coordination and communication. For each AI agent, the Coral server implements a specialized MCP server (the “Coral MCP server”) whose role is to facilitate agent-to-agent messaging through the notion of \textit{threads}. A thread is a structured conversation channel in which multiple agents (and potentially human users or other services) can participate. 
Coral MCP servers provide a suite of tools (accessible via SSE requests) that agents use to register and interact within this multi-agent environment. 
These tools include core operations for agent and thread management. Key tools exposed by the Coral MCP server are:

\begin{itemize}
\item \texttt{list\_agents}: Returns a list of all agents currently registered in the system, enabling discovery of potential collaborators or recipients of messages.
\item \texttt{create\_thread}: Creates a new communication thread (conversation) and specifies an initial set of participant agents. The thread functions as a context where messages are broadcast to all its participants. Each thread is identified by an ID, and its state (messages exchanged, participants list, etc.) is tracked on the Coral server.
\item \texttt{add\_participant} / \texttt{remove\_participant}: Modifies the participant list of an existing thread, allowing agents to join or leave ongoing conversations. This is useful for dynamically bringing in new expertise (via another agent) or removing agents that are no longer needed in the discussion.
\item \texttt{send\_message}: Transmits a message into a specified thread. The message typically contains content intended for one or more agents; it may include a direct mention (e.g., an @ tag or specific address) of a particular agent if a response from that agent is desired. When an agent invokes \texttt{send\_message}, the Coral server appends the message to the thread’s timeline and handles its propagation to participants.
\item \texttt{wait\_for\_mentions}: A blocking or subscription call that an agent uses to await any new messages in threads that mention the agent (i.e. explicitly address it). This mechanism is how an agent effectively “listens” for messages directed to it. Instead of the agent continuously polling for new data, the Coral server will notify it when relevant communication arrives.
\item \texttt{close\_thread}: Terminates a thread, optionally recording a summary or final outcome. Closing a thread signals that the conversation has concluded; resources can be freed and no further messages will be exchanged in that context.
\end{itemize}

These tools collectively realize a flexible communication protocol for the “Internet of Agents” – they allow agents to discover each other, form conversations, exchange information, and coordinate actions. 
Notably, the Coral server enforces structured messaging: messages are associated with threads and annotated with sender and (optionally) mentioned recipient metadata. This design ensures clarity in multi-party dialogues and allows the system to route messages appropriately (especially in directed queries or task delegations from one agent to another). The thread-based architecture also provides contextual compartmentalization: each conversation’s messages stay within its thread, preventing cross-talk and enabling per-thread context accumulation.

Internally, the Coral server's MCP implementation manages the state of agents and threads, and it ensures that notifications are dispatched when needed. For example, when AgentA sends a message in a thread mentioning AgentB, the server identifies that AgentB is referenced and generates an event to notify AgentB of the incoming message. This notification mechanism is central to how Coralised agents remain responsive to each other without continuous active querying.

In addition to tools, each Coral MCP server incorporates a dedicated cryptographic \textit{wallet} module.
Within Coral Protocol's design, this wallet underpins the trust and payment layer, enabling agents to engage in secure interactions.
In the current implementation, however, the wallet is used only to \textbf{receive} payments once a team task is successfully completed; agents do not yet initiate outbound transfers or other on‑chain operations.
\textit{The operational details of these payment flows are presented in Section \ref{sec:Secure-Payment-Mechanisms}.}

\subsubsection{MCP Servers}

The AI agent's capabilities can be augmented by using one or more (external) \textit{MCP servers}. An MCP server provides a standardized interface for the agent to access external tools, services, and secure identity functions. Essentially, an MCP server acts as an intermediary that the AI agent calls when it needs to perform actions beyond its native reasoning scope. 

An MCP server can host a suite of pluggable tools (e.g., external APIs, databases, computational utilities) that the agent can invoke in order to gather information or execute tasks. For instance, an agent might call a web search API, a code execution sandbox, or a weather service via these tool endpoints. By abstracting these capabilities behind a protocol, the agent can leverage them simply by issuing a tool-use request (often formulated as a special command or function call in the agent’s output) which the MCP server interprets and executes. This mechanism effectively extends the agent’s functionality with external context and operations, as is the goal of MCP – to enhance an AI model with external data and services in a consistent manner.

\subsection{Coralisation}

To transform external MCP servers or AI agents into fully functional Coralised Agents, Coral Protocol provides modular adapters known as \textbf{Coralisers}. 
A Coraliser acts as an onboarding layer, wrapping an external resource and exposing it via Coral Protocol using the Model Context Protocol (MCP).
This process is called Coralisation.
Coralisation allows previously siloed resources to participate in Coral-mediated complex tasks, with full interoperability, discoverability, and optional monetization.

Coralisation is categorized by the type of resource they integrate into the Coral ecosystem:

\begin{itemize}
    \item \textbf{MCP Coralisation} — Wrapping external MCP services or APIs into callable capabilities for Coralised Agents.
    \item \textbf{AI Agent Coralisation} — Onboarding legacy or third-party AI agents by mapping their interfaces to Coral standards. 
\end{itemize}

Each type is described in detail below.

\subsubsection{MCP Coralisation}

\textbf{MCP Coralisation} integrates external tools and APIs into the Coral ecosystem, enabling agents to invoke external services (e.g., simulations, transaction processors, APIs) as part of their reasoning loop. 
These tools can perform computations, automate real-world effects, or serve as actuators for AI decision-making.

\begin{figure}[ht!]
    \centering
    \includegraphics[width=\linewidth]{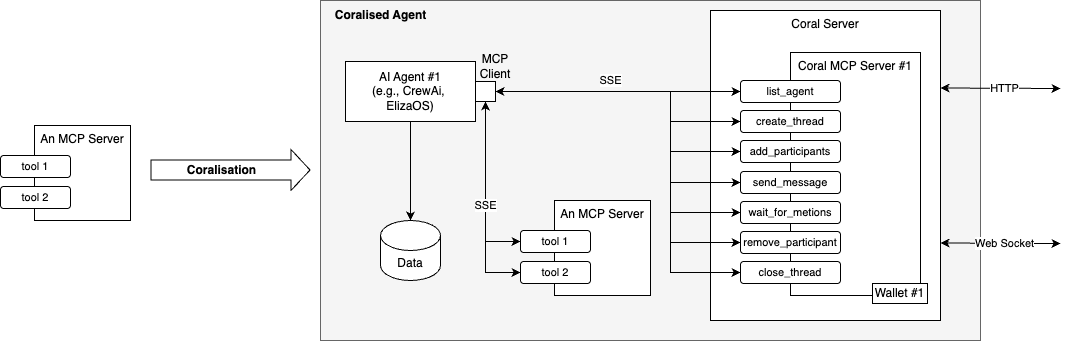}
    \caption{Coralisation of an MCP Server}
    \label{fig:Coralisation-of-MCP}
\end{figure}

MCP Coralisation wraps an outside MCP Server and translates its callable operations (e.g., \texttt{"simulate\_market"}, \texttt{"fetch\_weather"}, \texttt{"send\_email"}) into Coral's standard messaging format (e.g., HTTP, SSE\footnote{Server-sent events, \url{https://en.wikipedia.org/wiki/Server-sent\_events}.}). 
Return values are reformatted into Coral-compatible responses, optionally including metadata about execution status, latency, or confidence.
In effect, this process “coralises” a standalone MCP Server – turning it into a Coral-compatible agent that can communicate and collaborate with other agents in the ecosystem. 
Developers can achieve this simply by providing a configuration for the target MCP server and running the Coraliser utility, which automates the entire setup. The Coraliser generates ready-to-run Coral agent code for the external service, eliminating any need for custom wiring or complex integration code\footnote{Refer to the Coraliser GitHub repository (\url{https://github.com/Coral-Protocol/coraliser/}) for implementation details and examples, including the \texttt{coraliser\_settings.json} configuration and the Coraliser generation script.}.

Once an external service is coralised, it behaves like a native Coral agent. 
The Coraliser repository, for example, demonstrates how to coralise a web-scraping service (Firecrawl MCP) and a code assistant service (GitHub MCP) into Coral agents. 
The process to coralise an external MCP server typically involves the following steps:

\begin{itemize}
    \item \textbf{Configure the MCP endpoint}: Add the external server’s connection details (API endpoint, keys, etc.) to the Coraliser settings (e.g. updating the \texttt{coraliser\_settings.json} file for the target MCP).
    \item \textbf{Generate the Coral agent}: Run the Coraliser tool, which validates connectivity and automatically produces a new agent script for that MCP server. This script contains the logic for translating Coral protocol messages to the external service's API calls and vice versa.
    \item \textbf{Launch the agent in Coral}: Execute the generated agent script to register the external service as a Coral agent (for example, running python \texttt{firecrawl\_coral\_agent.py} for a Firecrawl MCP). Once launched, the coralised agent connects to the Coral Server and is ready to receive tasks from other agents, invoking the external service’s capabilities in response.
\end{itemize}

MCP Coralisers play a crucial role in Coral Protocol ecosystem by providing a seamless on-ramp for external AI functionalities. 
Any AI service that implements the MCP standard can be quickly onboarded as a trusted Coral agent. 
This dramatically extends Coral’s interoperability and capabilities: specialized model servers, tools, or third-party AI systems can all collaborate under Coral’s unified framework. 
By simplifying integration and automating adapter generation, MCP Coralisers make multi-agent systems more efficient, scalable, and production-ready without additional configuration overhead. 
They ensure that even outside models or services become first-class participants in Coral’s agent society, expanding the range of tasks and domains that Coral agents can collectively tackle.

\subsubsection{Agent Coralisation}

\textbf{Coralisation} can also onboard existing AI agents or multi-agent frameworks not originally designed for Coral. This includes agents with their own decision loops, internal memory, or planning logic. The Coraliser adapts the agent's messaging interface to Coral Protocol conventions, registers its capabilities, and mediates its participation in secure task executions.

\begin{figure}[ht!]
    \centering
    \includegraphics[width=\linewidth]{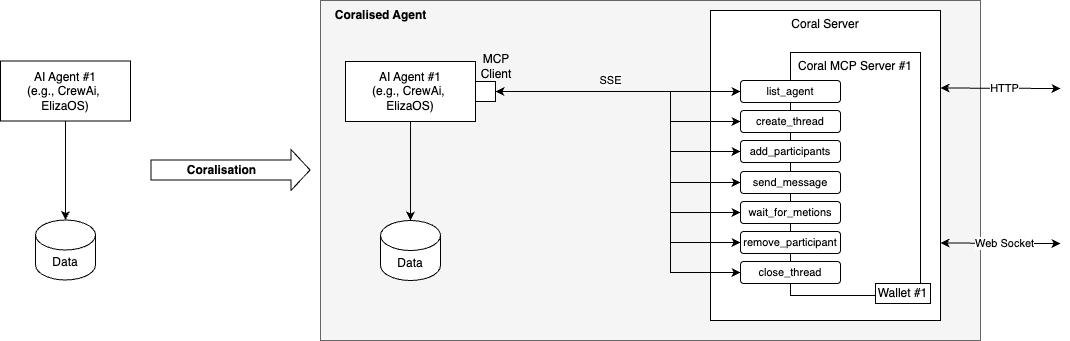}
    \caption{Coralisation of an AI Agent}
    \label{fig:Coralisation-of-AI-Agent}
\end{figure}

Coralisation of AI agents may encapsulate black-box agents from open-source projects, enterprise tools, or experimental research environments. By Coralising them, developers enable legacy agents to collaborate, receive payments, and participate in team formations alongside natively built Coralised Agents.

\subsection{Benefits of Coralisation}

Coralisers provide the foundation for a modular, scalable, and secure ecosystem. Key advantages include:

\begin{itemize}
  \item \textbf{Plug-and-Play Onboarding:} Any model, data, or agent can be Coralized without altering its internal logic.
  \item \textbf{Interoperability at Scale:} Coralisers enforce uniform schemas and interfaces, enabling smooth integration of heterogeneous resources.
  \item \textbf{Security and Governance:} Coralisers can attach access policies, rate limits, and usage conditions to integrated resources.
  \item \textbf{Economic Participation:} Each Coralised component can specify pricing and compensation terms, integrating into the ecosystem’s native token economy.
\end{itemize}

Together, Coralised Agents and Coralisation form the operational and extensibility backbone of Coral Protocol—enabling secure, multi-agent collaboration at Internet scale.

\newpage
\section{Secure Team Formation}

Many real-world tasks require the combined effort of multiple agents, each possessing distinct capabilities and credentials. To support such collaboration, Coral introduces a secure team formation mechanism that enables developers to assemble ad hoc coalitions of agents with verifiable identities, scoped permissions, and auditable trust scores.

Currently, the responsibility for orchestrating team composition, assigning roles, and coordinating tasks lies within the application layer. Developers manage the life-cycle of these teams manually; selecting agents, specifying their roles, and enforcing coordination logic through Coral’s secure communication and credentialing infrastructure.

Coral Protocol provides critical primitives to make such collaboration trustworthy:
\begin{itemize}
  \item \textbf{Identity and Trust Anchoring:} Each agent has a decentralized identifier (DID) and cryptographic credentials, allowing verifiable participation in multi-agent teams.
  \item \textbf{Team Contracts:} Multi-agent team formation can be codified via signed agreements, optionally stored on-chain, defining roles, responsibilities, and access policies.
  \item \textbf{Reputation-Based Selection:} Coral maintains a secure reputation mechanism where each agent's reputation score is stored on the blockchain. These scores reflect historical performance and trustworthiness, and are used to guide the selection of agents for new teams. Upon task completion, each agent’s reputation is updated based on performance metrics, peer evaluations, or objective task outcomes.
\end{itemize}

While task allocation and orchestration are currently managed manually by developers, future versions of Coral will offer a native Task Management service that can dynamically discover, coordinate, and supervise agent teams in response to complex user requests.

Further technical details and implementation specifics of these features will be provided in future versions of this whitepaper.

\newpage
%


\section{Secure Payment Mechanisms}
\label{sec:Secure-Payment-Mechanisms}

Coral’s Secure Payments system is implemented as an on-chain escrow program on the Solana blockchain, providing a trustless, token-based framework for agent transactions. 
This smart contract architecture manages payments in \textbf{SPL tokens} (Solana's token standard) and ensures funds are held in escrow until predefined conditions are met. 
By leveraging Solana's high-throughput, low-cost infrastructure, Coral can handle \textbf{fine-grained micropayments} and rapid fund releases with minimal fees. 
The result is a robust, decentralised payment layer where code-enforced rules guarantee that agents are paid \textbf{only when they fulfil agreed tasks}, and payers’ funds remain secure until those conditions are satisfied.

Coral is designed to support following multi-agent economic coordination:

\begin{itemize}
\item \textbf{Escrow Contracts (Live)}: For any multi-agent task, payments should be deposited into an escrow smart contract before work begins. The \textbf{Coral escrow program} holds the funds in a secure \textbf{vault account} under program control (see Algorithm \ref{alg:Coral-Escrow-Main-Program-Logic}). Funds remain locked until the contract’s release conditions are triggered – for example, when the requester signals task completion or an autonomous verification is recorded. At that point, the contract \textbf{automatically releases the payment} to the wallet of the fulfilment agent\footnote{i.e., the agent that fulfils the task and therefore gets paid.}. If the conditions are not met (e.g., the task fails or a deadline passes), the contract can refund the requester, eliminating counterparty risk without requiring any intermediaries. This conditional escrow mechanism is enforced entirely by smart contract logic, making the outcome \textbf{trustless} – neither party can unilaterally cheat or seize the funds.
\end{itemize}

All transactions in Coral’s payment system are recorded on Solana’s \textbf{immutable public ledger}, providing full \textbf{auditability} and accountability\footnote{Note that direct agent‑to‑agent payments are \textbf{intentionally not supported} within Coral Protocol.
A raw SPL‑token transfer sent from one wallet to another bypasses the escrow program’s signature checks, per‑agent caps, and six‑hour refund window, so any mistake, or deliberate fraud, becomes irreversible the moment it is confirmed on Solana.  Without the Session Vault acting as a cryptographic referee there is no on‑chain dispute mechanism, no automatic refund path, and no audit trail linking the payment to a specific task.  For that reason all compensation must flow through the coral‑escrow contract, where funds remain locked until the contract’s tamper‑proof rules are satisfied, protecting both payer and payee.}. 
Every deposit, partial release, and final disbursement is cryptographically signed and timestamped on-chain, allowing any participant to verify the history of payments for a task. 
This audit trail makes it possible to trace who paid whom, when, and under what conditions, forming a permanent record of service fulfillment. 
Moreover, because the escrow logic is executed by a public smart contract, the \textbf{payment conditions and outcomes are transparent}: agents cannot receive funds without meeting their obligations, and conversely, requesters cannot reclaim or redirect escrowed funds unless the contract’s rules (known and agreed upfront) allow it. 
The \textbf{smart contract enforces payments impartially}, removing the need for any trusted escrow agent or financial intermediary. 
This guarantees trustless collaboration – parties who may not know each other can confidently transact, knowing the blockchain will only release funds if the encoded conditions are satisfied.

To maximize usability and accessibility, Coral supports authentication abstraction. 
Users can log in using familiar methods (email, social accounts, enterprise SSO), with Coral managing wallet creation and key custody through provider-agnostic integrations with Wallet-as-a-Service (WaaS). 
Alternatively, Coral allows users who prefer full control to adopt a non-custodial approach by using their own wallets, enabling flexibility based on user preference and security requirements.
Coral's escrow contract establishes the cryptographic foundations essential for secure payments, but the protocol's capabilities extend significantly beyond simple fund locking.
Thanks to this authentication abstraction, Coral discreetly manages all blockchain operations behind the scenes. 
This clear separation between user experience and underlying blockchain mechanics is vital to driving widespread adoption, allowing Coral to bridge the gap between mainstream usability and secure decentralized payments.

\subsection{High‑level Intuition}
\label{sec:payment-intuition}

When a multi-agent application user (the \emph{authority}) starts work with a group of agents, she creates a \emph{Session Vault}:
a tiny Solana program‑derived account (PDA) that will
temporarily hold the budget for that task.
Each participating agent receives

\begin{enumerate}
\item a human‑readable \texttt{agent\_id} (e.g.\ \textit{``reviewer''}),
\item a destination SPL‑token address, and
\item an upper‑bound (\texttt{max\_cap}) on how much it may withdraw.
\end{enumerate}

Funds stay frozen until an agent proves, with a normal Ed25519 signature, that it is the rightful claimant.
If a claim is never made (e.g.\ the agent crashes), the
authority can safely \texttt{refund\_leftover()} after a six‑hour grace
period, so capital is never lost in limbo.

Natural terms such as “\emph{locked}’’, “\emph{claimed}’’ and “\emph{refunded}’’
map one‑to‑one to on‑chain state flags, giving non‑crypto stakeholders a
vocabulary that mirrors the ledger.

\begin{figure}[ht!]
    \centering
    \includegraphics[width=0.70\linewidth]{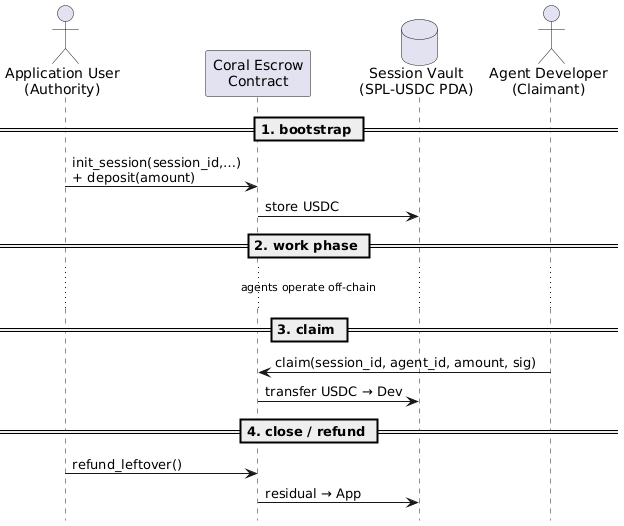}
    \caption{Lifecycle of a payment session, messages in \textcolor{darkgray}{grey}
           are optional.  The figure shows how funds flow \emph{once},
           eliminating re‑entrancy vectors.}
    \label{fig:Payment-Architecture}
\end{figure}

Figure~\ref{fig:Payment-Architecture} is the bird’s‑eye view: the application deposits
USDC, agents withdraw exactly once, and any residue returns to the
authority, no exotic multi‑hop escrow chains, just three deterministic
transitions.

\subsection{Contract Anatomy}

\begin{algorithm}[H]
\footnotesize
\caption{Coral Escrow Contract Logic}
\label{alg:Coral-Escrow-Main-Program-Logic}
\begin{algorithmic}[1]

\Procedure{InitSession}{$ctx$, $session\_id$, $operator$, $agent\_ids$, $payment\_wallets$, $developer\_pubkeys$, $max\_caps$}
    \State session.authority $\gets$ ctx.authority.key
    \State session.operator $\gets$ operator
    \State session.session\_id $\gets$ session\_id
    \State session.mint $\gets$ ctx.mint.key
    \State current\_time $\gets$ Clock.get().unix\_timestamp
        \State claim\_window $\gets$ DEFAULT\_CLAIM\_WINDOW\_SECONDS
    \State session.claim\_deadline $\gets$ current\_time + claim\_window
    \If{$length(agent\_ids) \neq length(payment\_wallets)$ \textbf{or} 
        $length(agent\_ids) \neq length(developer\_pubkeys)$ \textbf{or} 
        $length(agent\_ids) \neq length(max\_caps)$}
        \State \textbf{throw} InvalidInputLengthError
    \EndIf
    \If{$length(agent\_ids) = 0$}
        \State \textbf{throw} EmptyAgentsError
    \EndIf
    \If{$length(agent\_ids) > MAX\_AGENTS$}
        \State \textbf{throw} TooManyAgentsError
    \EndIf
    \If{ctx.escrow\_vault.mint $\neq$ ctx.mint.key}
        \State \textbf{throw} InvalidVaultMintError
    \EndIf
    \ForAll{cap \textbf{in} max\_caps}
        \If{cap $\leq$ 0}
            \State \textbf{throw} ZeroCapError
        \EndIf
        \If{cap < MIN\_CAP\_LAMPORTS}
            \State \textbf{throw} CapTooSmallError
        \EndIf
    \EndFor
    \State session.agent\_ids $\gets$ agent\_ids
    \State session.payment\_wallets $\gets$ payment\_wallets
    \State session.developer\_pubkeys $\gets$ developer\_pubkeys
    \State session.max\_caps $\gets$ max\_caps
    \State session.claimed $\gets$ array of size length(agent\_ids), initialized to \textbf{False}
    \State \textbf{return} Success
\EndProcedure

\Procedure{Deposit}{$ctx$, $session\_id$, $amount$}
    \If{amount $\leq$ 0}
        \State \textbf{throw} ZeroAmountError
    \EndIf
    \State Perform token transfer from depositor to escrow vault
    \State \textbf{return} Success
\EndProcedure

\end{algorithmic}
\end{algorithm}

Algorithm~\ref{alg:Coral-Escrow-Main-Program-Logic} illustrates the platform‑agnostic pseudo-code for the Coral escrow contract\footnote{In Solana's program architecture, several constraints such as account mutability and mint matching are enforced through instruction macros (e.g., \#[account(constraint = ...)]). For brevity and clarity, this section provides pseudocode that abstracts away these explicit macro-level constraints.}.
Three points deserve emphasis:

\begin{enumerate}
  \item \textbf{Single‑vault design.}\
        All liquidity for a session sits in \emph{one} token account,
        so every extra agent adds \(\mathcal{O}(1)\) keys, not new
        PDAs or rent overhead.
  \item \textbf{Bitmap bookkeeping.}\
        The \texttt{claimed[]} vector means replay attacks downgrade to
        a trivial double‑claim lookup.
  \item \textbf{\emph{Operator} role.}\
        Large installations can delegate refunds to an “ops’’ wallet
        without revealing the root authority keys.
\end{enumerate}

The Coral Escrow main program comprises two primary functions: \textbf{InitSession} and \textbf{Deposit}, each designed to securely manage payment sessions on the Solana blockchain.

\paragraph{InitSession Procedure.}
The \texttt{InitSession} procedure initializes and configures a new payment session. It first sets essential parameters such as the session's creator (\textit{authority}), an optional operational administrator (\textit{operator}), a unique \textit{session\_id}, and the type of tokens used (\textit{mint}). The procedure captures the current blockchain timestamp and sets a claim deadline based on whether the session is running in beta or standard mode.

It includes several validation checks:
\begin{itemize}
    \item Ensures arrays provided as input (agent IDs, payment wallets, developer keys, and payment caps) are of equal length.
    \item Verifies that at least one agent is specified and does not exceed a predefined maximum.
    \item Confirms the escrow vault uses the correct token type matching the session mint.
    \item Validates that each agent’s maximum payment cap is positive and above the allowed minimum.
\end{itemize}

After successful validation, the procedure records agent details, their associated wallets, public keys for developers, individual payment limits, and initializes a tracking mechanism for payment claims, marking them as unclaimed initially.

\paragraph{Deposit Procedure.}
The \texttt{Deposit} procedure manages the inflow of tokens into the session's escrow vault. 
It first validates that the deposited amount is positive.  
Once validation is complete, the tokens are transferred\footnote{Before transferring tokens from the depositor to the escrow vault, the contract verifies several constraints: it confirms the depositor’s token account and escrow vault share the same token type (mint) as defined by the session state, ensures the session state account matches the depositor’s authority, and validates all related account signatures and seeds to securely link accounts.} from the depositor’s account into the escrow vault, securing the funds until agents claim their payments.

Together, these procedures ensure a secure, consistent, and reliable mechanism for managing payments within Solana-based escrow sessions.

\subsection{How a real transaction feels}
\label{sec:payment-walkthrough}

To appreciate \emph{why} the escrow contract matters, it helps to walk through a
concrete end‑to‑end flow from four vantage‑points: (1) a non‑crypto product
manager, (2) an agent developer, (3) the application backend, and (4) a
compliance officer.  Figure~\ref{fig:Payment-Architecture}
shows the same life-cycle visually; the text below spells out the human
experience step‑by‑step.

\paragraph{1.~Product manager (no wallet, no jargon).}
\begin{enumerate}[label=\alph*)]
  \item Clara opens the SaaS dashboard, fills in a form saying  
        \textit{“Budget = 100 USDC, Claims close in 6h, Agents = reviewer,~tester.”}
  \item When she clicks \textsc{Start Session}, her credit‑card payment is
        auto–converted to USDC behind the scenes by the Wallet‑as‑a‑Service
        (WaaS) provider; a Solana transaction is signed for her with a custodial
        key and the \verb|init_session| + \verb|deposit| CPI sequence is
        broadcast.
  \item Clara receives an on‑screen URL
        \verb|https://explorer.solana.com/tx/5Y...| and a big green \textsc{LIVE}
        badge.  She never touches a seed‑phrase, yet funds are now locked in a
        publicly verifiable vault.
\end{enumerate}

\paragraph{2.~Agent developer (crypto‑native).}
\begin{enumerate}[label=\alph*)]
  \item Alex, who maintains the \texttt{reviewer} agent, has already whitelisted
        his programme‑derived wallet when publishing the agent to Coral.
  \item As soon as the agent finishes its static‑analysis job it calls  
        \texttt{claim(session\_id="20250707-42", agent\_id="reviewer", amount=40)}.  
        The call is wrapped in Anchor’s \verb|#[derive(Accounts)]| struct and
        forwarded by a lightweight TypeScript relayer.
  \item The contract verifies his Ed25519 sig in micro‑seconds and emits a
        \texttt{Claimed{…}} log.  Alex can see the token balance in his Phantom
        wallet within ~400 ms (one Solana confirmation).
\end{enumerate}

\paragraph{3.~Backend / authority.}
\begin{enumerate}[label=\alph*)]
  \item A webhook on the application server listens for the
        \texttt{Claimed} event; it marks the “review step” as \textsc{Done} in
        Postgres and notifies Slack.  No polling, no race‑conditions.
  \item Six hours after session start, a cronjob executes
        \verb|refund_leftover()|.  
        Any unclaimed USDC flows back to Clara’s treasury wallet in the same Tx,
        closing the PDA and freeing rent.
\end{enumerate}

\paragraph{4.~Compliance / finance.}
\begin{enumerate}[label=\alph*)]
  \item Finance downloads a CSV from the Solana Explorer API once a month and
        reconciles payments by matching the memo field
        (\texttt{"session:20250707-42"}).  
        Because every transfer is a L1 token‐transfer, totals can be audited
        independently of the Coral database.
  \item The refundable six‑hour window, the per‑agent \texttt{max\_caps}, and
        the contract’s overflow‑safe maths satisfy the company's segregation‑of‑duty
        and risk limits, nothing to re‑implement off‑chain.
\end{enumerate}

\bigskip
\noindent
\textbf{Why this matters.}  
Users who are skeptical of blockchains can benefit from seamless experiences without being exposed to the underlying technology. 
Meanwhile, advanced users retain the ability to inspect every transaction on a transparent public ledger. Businesses gain the advantage of provable settlement finality within seconds, rather than waiting for days or weeks.
Algorithm~\ref{alg:Coral-Escrow-Main-Program-Logic} illustrates the minimal code required to achieve this functionality, just four public instructions and fewer than 250 logical lines.

\subsection{Road‑map}
\label{sec:Payment-Road‑map}

Smart‑contract payment rails must balance \emph{safety}, \emph{usability}, and
\emph{feature velocity}. Shipping everything at once would slow audits to a crawl; shipping nothing erodes developer trust. Coral thus follows an incremental four-stage roadmap, each with tight, security-reviewed scope and a clear “go/no-go” criterion.

\begin{center}
\begin{tabular}{@{}p{4.0cm}p{1.8cm}p{5.0cm}p{2.2cm}@{}}
\toprule
\textbf{Stage} & 
\textbf{Date} &
\textbf{Feature} & \textbf{Status} \\\midrule
1 (Crypto Native) & Main-net & Direct deposits, withdrawals, full custody & \emph{Live} \\
2 (Mainstream Bridge) & Q3 2025 & Fiat on-ramp, invisible wallets, gasless tx & In progress \\
3 (Trust Enhancement) & Q4 2025 & Reputation-driven agent selection, quality scoring & Planned \\
4 (Economic Security) & 2026 & Agent staking, slashing, decentralized QA & Research \\
\bottomrule
\label{tab:payment-roadmap}
\end{tabular}
\end{center}

Table~\ref{tab:payment-roadmap} illustrates the quarterly rollout cadence, allowing a four-week buffer for third-party audits and backward-compatibility testing after each main-net deployment.

\bigskip
\noindent\textbf{Stage 1: Crypto Native.}
Algorithm~\ref{alg:Coral-Escrow-Main-Program-Logic} already supports direct token deposits, withdrawals, and refunds, providing full user custody and transparent on-chain operations. This baseline stage prioritizes \emph{invariant safety} (caps, deadlines, signature checks). A two-week public beta on Solana \texttt{testnet} processed \$73k in mock USDC without invariant breaches, enabling immediate main-net deployment.

\medskip
\noindent\textbf{Stage 2: Mainstream Bridge.}
Developer feedback underscored the need for broader accessibility:
\begin{enumerate}
\item Fiat payments via credit cards (automatic USDC conversion).
\item Invisible wallet management (Wallet-as-a-Service).
\item Gasless transactions (relayer-signed variant of \texttt{claim}).
\item Email-based authentication for seamless onboarding.
\end{enumerate}
As these upgrades involve minimal changes to critical security logic, only delta audits are required, streamlining review.

\medskip
\noindent\textbf{Stage 3: Trust Enhancement.}
Coral introduces a reputation-driven selection process, leveraging a quality scoring system to match agents to tasks optimally. This stage includes:
\begin{enumerate}
\item Continuous reputation indexing.
\item Quality filtering for agent selection.
\item Performance-based dynamic pricing.
\end{enumerate}
A lightweight oracle maintains real-time health checks to prevent overdrafts or service interruptions.

\medskip
\noindent\textbf{Stage 4: Economic Security.}
The final stage enforces economic safeguards to enhance decentralized quality assurance, including:
\begin{enumerate}
\item Agent staking to ensure commitment.
\item Slashing mechanisms for penalizing underperformance.
\item Fully decentralized quality assurance driven by autonomous smart contracts.
\end{enumerate}
Formal verification of economic incentive structures and denial-of-service safeguards is complex, thus planned for thorough research and prototyping in 2026.

\paragraph{Risk management and audits.}
Each stage includes:
\begin{itemize}
\item \textbf{Static analysis}: MIR-level overflow scanning, Anchor linting.
\item \textbf{Property-based fuzzing}: 500 randomized runs per CI build.
\item \textbf{External audit}: Sec3, OtterSec, Neodyme rotation.
\item \textbf{Bug bounty}: 30-day Immunefi period before main-net deployment.
\end{itemize}

Progression is strictly contingent on addressing all audit findings, safeguarding the integrity and robustness of the core system as illustrated in Figure~\ref{fig:Payment-Architecture}.

\bigskip
\noindent\emph{Bottom line:} This roadmap preserves rapid developer iteration without compromising on-chain security, progressively introducing advanced payment functionalities in an audit-friendly and backward-compatible manner.

\subsection{Take‑away}


The on-chain Secure Payments architecture transforms Coral into a decentralized marketplace of agent services. 
Agents can \textbf{price their services transparently} and even receive streaming or micro-payments for incremental work, knowing that the funds will be delivered automatically when they perform as agreed. 
Requesters, in turn, gain confidence that their payments will only be released for verifiable outcomes, and can retrieve funds if work is not delivered. 
The \textbf{entire payment life-cycle is governed by code}, not by trust in a third party. In essence, Coral’s payment layer provides an incentive-aligned economic foundation for the Internet of Agents: it aligns agent rewards with successful task completion, supports rich collaboration models (from one-to-one gigs to multi-agent revenue sharing), and maintains a tamper-proof ledger of all economic exchanges. 
This secure economic infrastructure underpins a network of commercially autonomous agents that can collaborate and transact freely, forming a self-sustaining agent economy built on transparency and trustless guarantees.

\newpage
\section{Exploiting Coral Protocol to a Real-World Application}
\label{sec:Exploiting-Coral-Protocol-to-a-Real-World-Application}

\subsection{Envisioned Coral Ecosystem}

In this section, we illustrate how Coral Protocol can successfully be exploited as a design in a real world application.

\begin{figure}[ht!]
    \centering
    \includegraphics[width=1.0\linewidth]{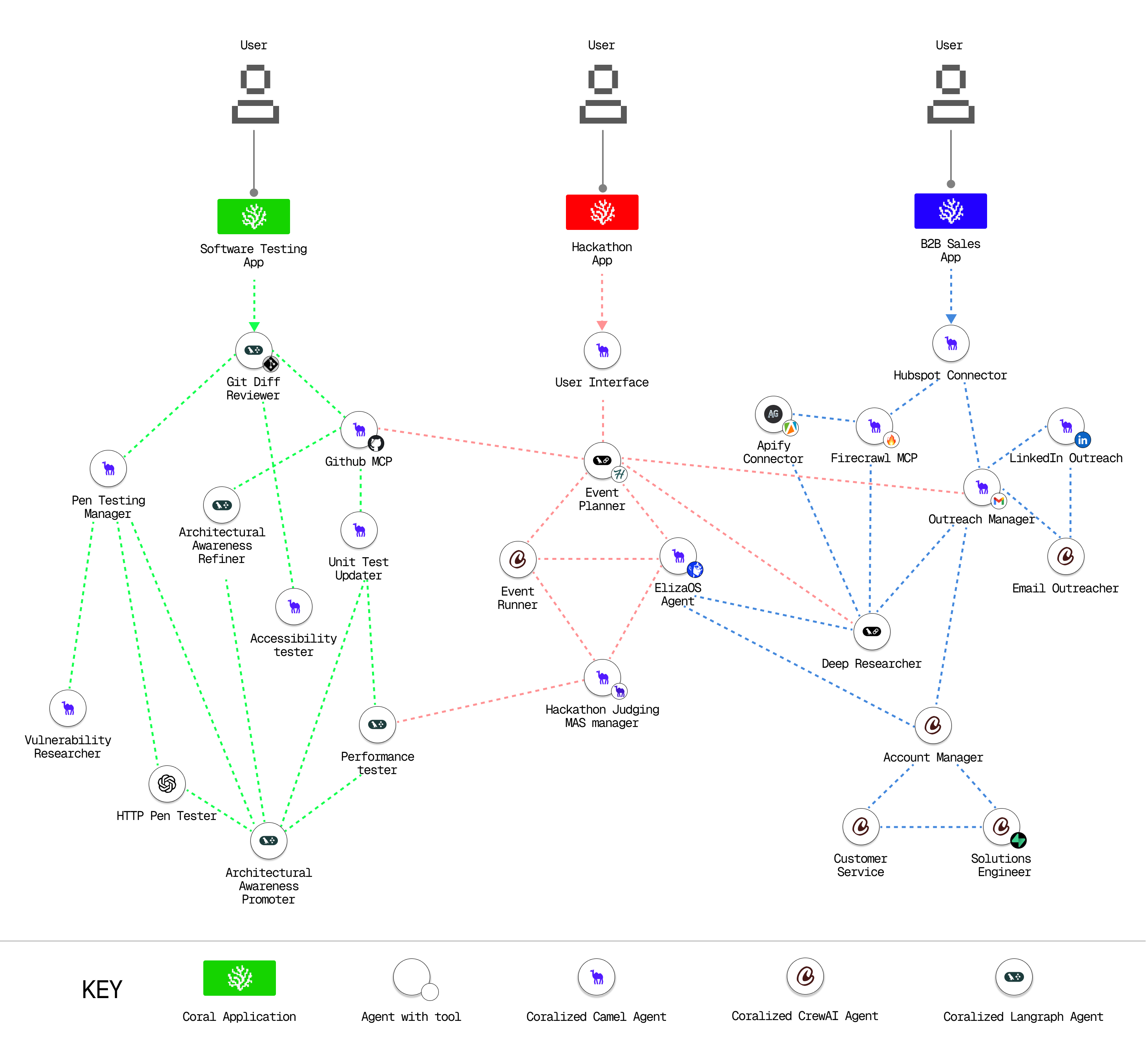}
    \caption{The Coral Use Case Example}
    \label{fig:Coral-Use-Case}
\end{figure}

Consider the diagram illustrated in Figure \ref{fig:Coral-Use-Case}. 
The diagram shows a mesh of reusable micro-agents, all speaking Coral Protocol, assembled into \textbf{three applications} whose components (which are \textit{Coralized} agents) communicate exclusively through the \emph{Coral} protocol.  
Every dotted edge is a Coral channel; an agent that terminates several colours is shared by (and context-switches between) products without glue code.
Three coloured sub-meshes, each for one multi-agent application (blue for the B2B‐Sales product, red for the Hackathon product, green for the Software-Testing product), coexist on the same network:

\begin{itemize}
    \item The B2B sales product is boot-strapped by HubSpot web-hooks that open a Coral session as soon as a new lead appears in the customer’s CRM.  A \emph{HubSpot agent} collects the lead record and its pipeline context, then cooperates with a \emph{Firecrawl MCP agent}—generated automatically from a Firecrawl server by the Coralizer—to enrich the prospect with publicly available data.  Once the profile is complete, the agent hands control to an \emph{Outreach Manager}, which coordinates multichannel nurturing campaigns.  Whenever additional insight is required, follow-up queries are routed to a \emph{Deep Research agent}; that agent, in turn, consults an \emph{ElizaOS agent} both to harvest the latest social-media signals and, when appropriate, to publish celebratory community posts for leads that have converted to partnerships.  Throughout the process, every step is expressed as Coral envelopes, so progress and state updates flow back into HubSpot without custom glue code.

    \item The hackathon product offers organisers a chat-centric control room.  A dedicated \emph{User Interaction agent} receives commands from the human organiser and relays them to an \emph{Event Planner agent}.  The planner calls on the \emph{Deep Research agent} to ground every decision in the chosen event theme and, when a sponsoring repository is involved, leverages a \emph{GitHub MCP agent} produced by the Coralizer to pull project metadata directly from GitHub.  In the run-up to the event, an \emph{ElizaOS agent} promotes sign-ups across social channels; on the day itself, an \emph{Event Runner agent} orchestrates real-time logistics while liaising with a coralised, multi-agent judging system to select the winners.  Judges augment their assessment with metrics from a \emph{Performance Testing agent}, ensuring that the final ranking reflects both creativity and technical quality—all without leaving the Coral mesh.

    \item Whenever a new commit lands on the user’s repository, GitHub emits a webhook that the Software Testing application translates into a Coral session.  It forwards the commit hash to a \emph{Git Diff Review agent}, which checks out the code and analyses the delta.  To understand the broader context, the reviewer queries the \emph{GitHub MCP agent} for the most relevant linked issue or pull request.  Findings are then cross-validated through a trio of specialist agents: the \emph{Performance Testing agent}, the \emph{Pentesting Management agent} (which itself orchestrates deeper security probes such as an HTTP pen-tester), and the \emph{Accessibility Testing agent}.  Only when all three corroborate the review does the change progress; otherwise, annotated feedback is pushed back to the developer via GitHub checks.  Because every interaction rides over Coral, quality gates can be added, removed, or replaced at will, and the entire pipeline remains resilient to individual agent failures.

\end{itemize}


The deployment captured in Figure \ref{fig:Coral-Use-Case} demonstrates that \textbf{Coral is more than an interface specification—it is an architectural substrate}.
By enforcing envelope-level contracts and embracing message-centric composition, Coral enables:

\begin{itemize}
    \item \textbf{Effortless reuse}. The same micro-agent can serve several products concurrently without code forks.
    \item \textbf{Incremental evolution}. Agents can be versioned, hot-swapped, or rolled back in isolation, as long as they honour their declared Coral schemas.
    \item \textbf{Polyglot freedom}. Teams author agents in the language, framework, or runtime that best suits their domain; Coral guarantees wire-level interoperability.
    \item \textbf{Operational resilience}. Network partitions or agent crashes localise failure, while typed \textbf{N}egative \textbf{ACK}nowledgements propagate intent-aware fall-back paths.
    \item \textbf{Faster time-to-value}. The Coralizer turns any well-formed API into a drop-in agent, shrinking the integration backlog from weeks to minutes.
\end{itemize}

In short, the mesh validates Coral’s promise: \emph{a protocol that lets small, purpose-built micro-agents snap together like LEGO bricks, yet scale to support entire product lines}. What begins as three discrete applications quickly converges into a living ecosystem where capabilities are shared, not rewritten—turning integration effort into a strategic asset rather than a sunk cost.

\subsection{Intelligent Software Testing Application}

To demonstrate the real-world capabilities of Coral Protocol, we developed a multiframework intelligent software testing application that orchestrates specialized AI agents using GitHub commits as a trigger point. 
The example, located in the Coral repository\footnote{\url{https://github.com/Coral-Protocol/coraliser/tree/multi-frameworks/coral_examples/multiframework-github-testing}, accessed on May 17, 2025.}, showcases Coral Protocol's ability to dynamically compose agents across frameworks into coordinated tasks with minimal integration effort.

In this application, a developer commits code to a GitHub repository. This action triggers the Coral GitHub Coraliser, which captures the commit hash and initiates a multi-agent testing complex task.
The ... scenario file defines the flow.

\newpage
\section{Conclusion}

In summary, Coral Protocol provides a standardized framework for effective 
collaboration among specialized AI agents. By defining clear communication 
patterns and tools, it directly addresses the coordination challenges inherent 
in multi-agent systems while preserving the advantages of a modular, specialized
 design. As AI systems grow more complex and specialized, frameworks like the 
Coral Protocol will become increasingly crucial for building coherent solutions 
that leverage diverse capabilities to solve complex problems.

Beyond its technical merits, Coral Protocol carries significant practical 
and strategic potential. Widespread adoption of a common agent communication 
standard can foster an ecosystem in which any compliant agent—whether developed 
by a large organization or a small team—can seamlessly integrate and cooperate 
with others. This interoperability could enable network effects, where the 
addition of new agents increases the overall value and capability of the system 
for everyone. In an analogy, Coral aims to be like a universal “language” or 
interface for AI agents, allowing them to plug into collaborations as easily as 
devices connecting via a common port.

\medskip

\noindent\textit{This white paper presents the conceptual framework of the Coral
 Protocol for multi-agent collaboration. Implementation details and technical 
specifications may evolve as the protocol matures. We invite feedback and 
collaboration from the community as we refine this standard for the benefit of 
all stakeholders in the AI ecosystem.}

\newpage
\bibliographystyle{splncs04}
\bibliography{references}

\end{document}